\documentclass[a4paper,11pt]{article}
\usepackage{jinstpub,amsmath,amssymb,braket,bm,booktabs,color,nicefrac,url,verbatim, subcaption}
\usepackage[page]{appendix}  
\RequirePackage{doi}

\usepackage{ragged2e}
\DeclareCaptionJustification{justified}{\justifying}
\newcommand{\nikhef}[1]{\author[a]{#1}}
\affiliation[a]{Nikhef and the University of Amsterdam, Science Park, 1098 XG, Amsterdam, The Netherlands}

\newcommand{\zurich}[1]{\author[b]{#1}}
\affiliation[b]{Physik-Institut, Universit\"{a}t Z\"{u}rich, 8057~Zurich, Switzerland}

\newcommand{\purdue}[1]{\author[c]{#1}}
\affiliation[c]{Department of Physics and Astronomy, Purdue University, West Lafayette, IN 47907, USA}

\newcommand{\michigan}[1]{\author[d]{#1}}
\affiliation[d]{Michigan Ion Beam Laboratory, University of Michigan, Ann Arbor, MI 48109, USA}

\newcommand{\cbpf}[1]{\author[e]{#1}}
\affiliation[e]{Centro Brasileiro de Pesquisas F\'{i}sicas - CBPF, R. Dr. Xavier Sigaud, 150 - Urca, Rio de Janeiro, Brazil}

\newcommand{\freiburg}[1]{\author[f]{#1}}
\affiliation[f]{Physikalisches Institut, Universit\"{a}t Freiburg, 79104 Freiburg, Germany}

\newcommand{\asu}[1]{\author[g]{#1}}
\affiliation[g]{Simon A. Levin Mathematical, Computational and Modeling Sciences Center, Arizona State University, Tempe, AZ 85287, USA}

\newcommand{\1}[1]{\, \mathrm{#1}} 

\newcommand{\cs}{${}^{137}\mathrm{Cs }$}
\newcommand{\mn}{${}^{54}\mathrm{Mn }$}
\newcommand{\ti}{${}^{44}\mathrm{Ti }$}
\newcommand{\ba}{${}^{133}\mathrm{Ba }$} 
\newcommand{\co}{${}^{60}\mathrm{Co }$}
\newcommand{\pot}{${}^{40}\mathrm{K }$}
\newcommand{\gam}{$\gamma$}
\newcommand{\scinot}[1]{$\times 10^{#1}$}
\newcommand{\radunits}{Bq/$\mathrm{m}^{3}$}
\newcommand{\order}[1]{$\mathcal{O}(10^{#1})$}
\newcommand{\cel}[1]{$#1^{\circ}\mathrm{C}$}
\newcommand{\degree}[1]{$#1^{\circ}$}
\begin{document}

\title{A Precision Experiment to Investigate Long-Lived Radioactive Decays}

\nikhef{J.~R.~Angevaare}
\zurich{P.~Barrow}
\zurich{L.~Baudis}
\nikhef{P.~A.~Breur}
\zurich{A.~Brown}
\nikhef{A.~P.~Colijn}
\emailAdd{a.p.colijn@nikhef.nl}
\purdue{G.~Cox}
\zurich{M.~Gienal}
\nikhef{F.~Gjaltema}
\purdue{A.~Helmling-Cornell}
\purdue{M.~Jones}
\zurich{A.~Kish}
\purdue{M.~Kurz}
\michigan{T.~Kubley}
\purdue{R.~F.~Lang}
\emailAdd{rafael@purdue.edu}
\cbpf{A.~Massafferri} 
\cbpf{R.~Perci} 
\purdue{C.~Reuter}
\emailAdd{cassiereuter@purdue.edu}
\nikhef{D.~Schenk}
\freiburg{M.~Schumann}
\asu{S.~Towers}

\maketitle{}

\begin{abstract}

Radioactivity is understood to be described by a Poisson process, yet some measurements of nuclear decays appear to exhibit unexpected variations. Generally, the isotopes reporting these variations have long half lives, which are plagued by large measurement uncertainties. In addition to these inherent problems, there are some reports of time-dependent decay rates and even claims of exotic neutrino-induced variations. We present a dedicated experiment for the stable long-term measurement of gamma emissions resulting from $\beta$ decays, which will provide high-quality data and allow for the identification of potential systematic influences. Radioactive isotopes are monitored redundantly by thirty-two $76\1{mm}\times76\1{mm}$ NaI(Tl) detectors in four separate temperature-controlled setups across three continents. In each setup, the monitoring of environmental and operational conditions facilitates correlation studies. The deadtime-free performance of the data acquisition system is monitored by LED pulsers. Digitized photomultiplier waveforms of all events are recorded individually, enabling a study of time-dependent effects spanning microseconds to years, using both time-binned and unbinned analyses. We characterize the experiment's stability and show that the relevant systematics are accounted for, enabling precise measurements of effects at levels well below \order{-4}. 

\end{abstract}

\section{Introduction}
The decay of radioactive isotopes is regarded as a well-understood Poisson process, independent of external influences~\cite{Rutherford1910}. Early experiments tried to affect decay rates by varying the temperature between $24\1{K}$~and~$1280\1{K}$, by applying pressure up to $2000\1{atm}$, and by increasing the magnetic field up to $83\1{kG}$~\cite{meyer1927, kohlrausch1928, bothe1933}. All of these experiments either gave null results, or showed effects that could be explained with changes in the counting geometry. There are a few cases where minor changes in decay rates have been measured due to artificially produced changes in the physical environment of the decaying nuclides~\cite{Goodwin:2009,Emery:1972,Hopke:1974,Hahn:1976,Dostal:1977,Liu:2000,Norman:2001}. However, these changes occur under extreme conditions only, such that the structure of the isotope's electron cloud is deformed, and cause very small changes to the decay rate. As an example, pressures $\geq$ 400~kbar cause only a $1\%$ change in decay rate~\cite{Liu:2000}.

In recent years, however, a few independent groups have claimed to observe variations in nuclear decay constants of several long-lived $\alpha$, $\beta$ and \gam~-emitting isotopes~\cite{Ellis1990, Shnoll1998, Baurov2007, Parkhomov2010A, Parkhomov2010B, Baurov:2013}. These claims generally report an annual variation at the level of \order{-3}. A summary of these reports can be found in~\cite{Jenkins2012B}. One particular isotope that has received much recent attention is \ti{}, the decay rate of which has seen both claims~\cite{O'Keefe:2012} and counter-claims~\cite{Norman2009} of annual modulation. Another particularly well-studied example comes from the DAMA/LIBRA experiment, which aims to detect dark matter-induced scatters using NaI(Tl) detectors~\cite{Bernabei2008}. The collaboration observes a sinusoidal variation in the background rate at 9.3~$\sigma$ significance~\cite{Bernabei2005,Bernabei2013}, and claims this to be evidence for an annually modulated dark matter signal. However, all proposed dark matter models have been ruled out by other experiments~\cite{Agnese:2014, Aprile2015, Akerib2017, Amole2017, Aprile:2017B}. Curiously, the observed modulation coincides with the energy of X-ray emissions from the $^{40}$Ar electron capture process~\cite{Bernabei2008, Pradler:2012A, Pradler:2012B}. This lack of an accepted explanation for DAMA/LIBRA modulation warrants further investigation into systematic sources that can impact counting experiments using NaI(Tl) detectors. 

While most of the isotopes presented in~\cite{Jenkins2012B} show variations with periods of a year, \cs{} has received scrutiny because of its reported daily variations~\cite{Baurov2007}. Experimental evidence~\cite{Bellotti2012} excludes variations larger than $8.5$\scinot{-5}, for \cs{}, which were first reported in~\cite{Baurov2001} and attributed to solar influence in~\cite{Fischbach2009}. There have also been reports of changes in the activity of \mn{} correlated with solar activity~\cite{Jenkins2008}. Though these claims have since been excluded~\cite{Bellotti2013}. It is more likely that the variations seen in~\cite{Baurov2007} and~\cite{Jenkins2008}, among others~\cite{Ellis1990, Shnoll1998, Parkhomov2010A, Parkhomov2010B, Baurov:2013} are due to some local systematic effect, which has not yet been taken into account (see e.g.~\cite{Schrader2016}). There are some correlations between the ambient radon concentrations and the decay rates of alpha emitters~\cite{Pomme2017alpha}, but no consistent variations have been found with beta-emitters and electron capture decays~\cite{Pomme2017betaec}. Interestingly, variations that have been observed in one geographic location do not appear in another~\cite{Pomme2017betaec}.

Taken together, the data point to the existence of yet-unidentified systematic uncertainties in half life measurements, on the order of years, which we refer to as long-lived. We present an experiment that is designed to identify systematic effects responsible for such uncertainties. In particular, we present a set of four independent NaI(Tl)-based setups that monitor several isotopes in different geographic locations. Systematic effects are accounted for by geographical distance, including the seasonal offset from the Northern/Southern hemispheres. We employ a data acquisition system that records the waveform of each individual event which allows studies of time-dependent effects with microsecond precision. Various environmental and operational parameters, including background radiation, are monitored for detailed correlation studies. In addition to the identification of sources of systematic uncertainty in the measurement of long-lived radioactive decays, our experiment allows for precise measurements of the half lives of long-lived radioisotopes to improve on the literature values for selected isotopes.  Given the expected number of statistics, we will be able to make competitive measurements for half life within two years of data taking. 

\section{Experimental Setup}
The experiment consists of four identical setups located at Purdue University in the U.S., Nikhef in the Netherlands, the University of Z\"{u}rich in Switzerland and the Centro Brasileiro de Pesquisas F\'{i}sicas (CBPF) in Brazil, detailed in Tab.~\ref{table:locations}. Each setup houses eight NaI(Tl) detectors, arranged in pairs. Three of the detector pairs each measure a radioactive 1"~disc source. The remaining detector pair monitors the radioactive background, serving as a control for any ambient radiation present in the laboratory. Each of the isotopes are monitored in at least two locations. Currently studied isotopes are the commercially available \mn{}, \co{}, \ba{} and \cs{}, along with a \ti{} source manufactured at the Paul Scherrer Institute in Switzerland. The sources, their half lives, and locations are all detailed in Tab.~\ref{table:sources}. All sources were acquired with initial activities of $\sim 1\1{kBq}$, thus providing sufficient statistics while keeping pile up of events in the waveforms to a minimum.

\begin{table}[!htbp] \footnotesize
\begin{center}
\begin{tabular}{ l l l l }
\hline
Institute & Latitude & Longitude & Altitude [m] \\
\hline 
Purdue	& 40.4259\degree{} N	&	86.9081\degree{} W	&	187	\\
Nikhef 	& 52.3564\degree{} N	&	4.9508\degree{} E	&	-3	\\
Zurich	& 47.3743\degree{} N	&	8.5510\degree{} E	&	500	\\
CBPF	& 22.9544\degree{} S	&	43.1676\degree{} W	&	16	\\
\end{tabular}
\end{center}
\caption{The four locations where the setups are housed.  Purdue is located in West Lafayette, Indiana, in the United States.  Nikhef is the Dutch National Institute for Subatomic Physics, located in Amsterdam in the Netherlands.  Also in Europe is the University of Zurich in Zurich, Switzerland.  The different geographic locations enable laboratory systematics to be decoupled and allows daily modulations to be probed.  Finally, one setup is at the Centro Brasileiro de Pesquisas F\'{i}sicas (CBPF), in Rio de Janeiro, Brazil.  Because this setup is in the southern hemisphere, seasonal effects can also be decoupled.}
\label{table:locations}
\end{table}

\begin{table}[!htbp] \footnotesize
\begin{center}
\begin{tabular}{ r r l l l l l l l }
\hline
Source& Half Life & Ref. & \multicolumn{4}{c}{Institute} \\
\hline 
\ti 	& $59.1(3)\1{y}$ 		&~\cite{Chen:2011} 	& CBPF	&Nikhef	&Purdue	&Zurich \\ 
\mn 	& $312.2(2)\1{d}$		&~\cite{dong2014}		& 		&		&Purdue	&Zurich\\
\co 	& $5.271(4)\1{y}$ 	&~\cite{Tuli:2003}		& CBPF	&Nikhef	&Purdue	&Zurich\\ 
\cs 	& $30.08(9)\1{y}$ &~\cite{Browne:2007}		& CBPF	&Nikhef	&&\\ 
\end{tabular}
\end{center}
\caption{Radioactive isotopes currently monitored in the experiment. Each isotope is monitored in at least two locations, and these are the initial configurations.}
\label{table:sources}
\end{table}

\begin{figure}[!htbp]
\captionsetup[sub]{font = large, labelfont={bf, sf}}
\begin{subfigure}{0.45\textwidth}
\begin{center}
\includegraphics[width=\columnwidth]{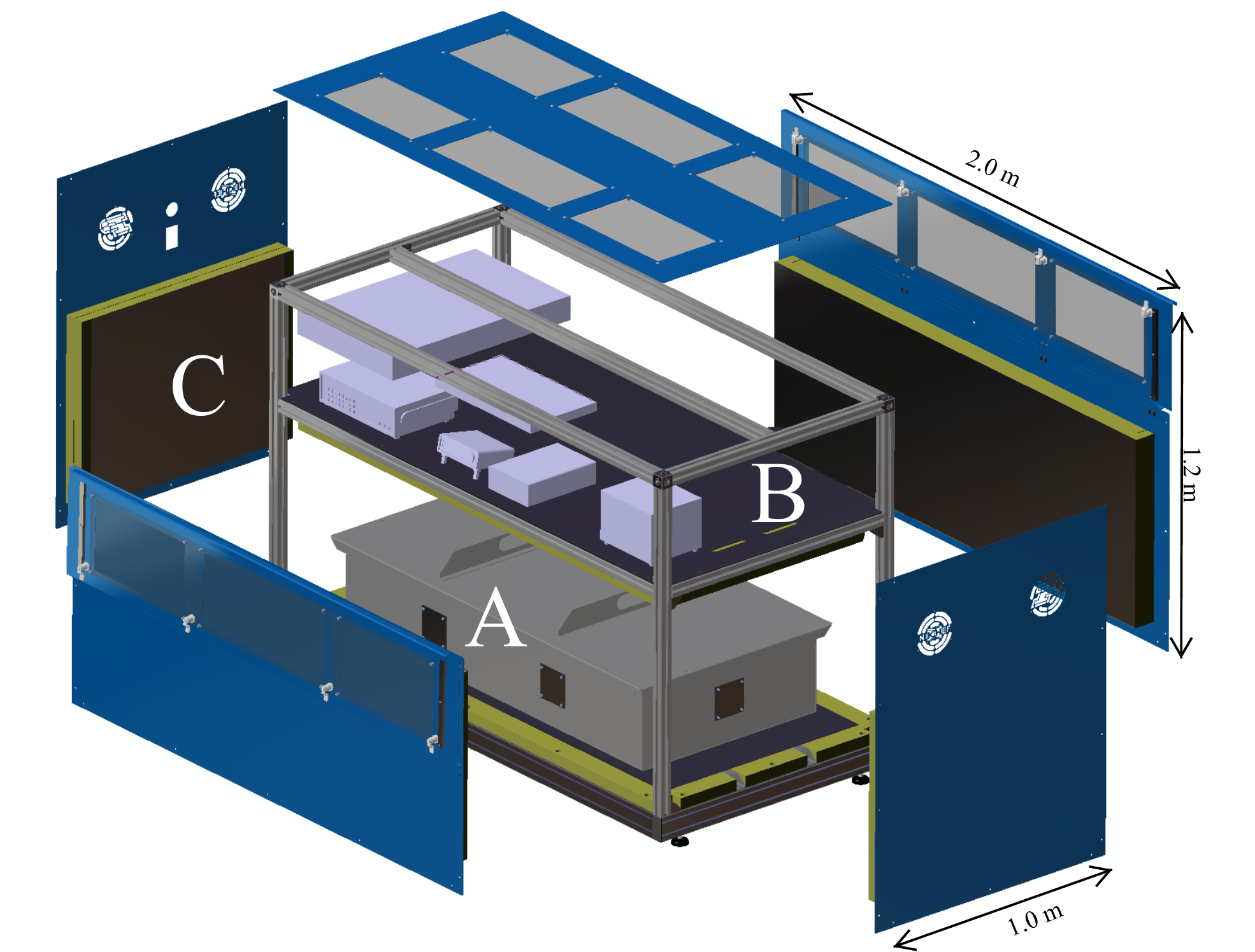}
\end{center}
\caption{\label{fig:box}}
\vspace{0.001\textwidth}
\end{subfigure}
\begin{subfigure}{0.6\textwidth}
\begin{center}
\includegraphics[width=\columnwidth]{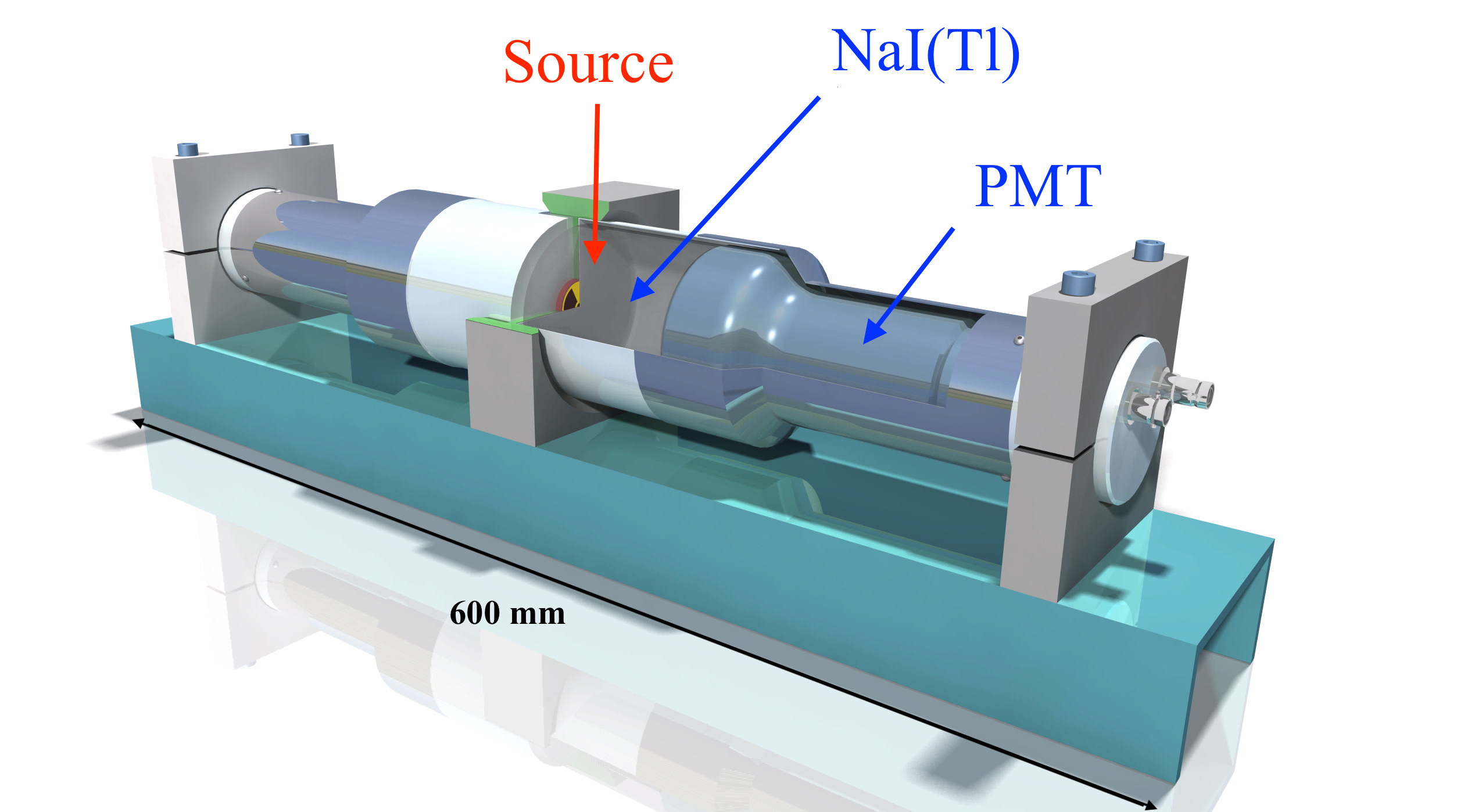}
\end{center}
\vspace{-0.01\textwidth}
\caption{\label{fig:detectorset}}
\vspace{-0.02\textwidth}
\end{subfigure}

\caption{Left: An exploded view of one setup. The inner box (A) on the lower rack contains the sources, detectors, lead shielding, heaters, and environmental sensors. The inner box is thermally insulated by a $10\1{cm}$ thick layer of polyurethane foam (C) and temperature stabilized. The air-cooled top rack houses the electronics (B) such as high voltage supplies, radon monitor, data acquisition system and host computer.  Right: Rendering of the geometry of a detector pair. An aluminum bracket holds the source between the two detectors. This allows for almost $4\pi$ coverage. The right detector is cut away and the NaI(Tl) crystal is not shown in order to reveal the positioning of the source.}
\end{figure}

A single setup is depicted in Fig.~\ref{fig:box}, with the sources being mounted in the innermost of the two boxes, located on the lower rack. In this inner box, the temperature is controlled through the use of heaters, and a variety of environmental parameters are monitored. It can be flushed with nitrogen to reduce the presence of $^{222}$Rn. Each source is sandwiched between two $3"\times3"$~NaI(Tl) detectors as in Fig.~\ref{fig:detectorset}. The distance between the two detector windows is $6.35\1{mm}$. Each of these four detector pairs are shielded by $5 \1{cm}$ of lead and separated by $\sim10\1{cm}$ of distance.

Taking into account the detector materials and geometry, we use Geant4~\cite{Agostinelli:2002} to simulate a detection efficiency for each detector pair of $1.4\%$, $11.3\%$, $22\%$, $6.2\%$ and $5.4\%$ for \pot{}, \cs{}, \mn{} and the $1.1\1{MeV}$ and $1.3\1{MeV}$ \co{} photopeaks, respectively. Though \pot{} is not explicitly measured, it is a large source of background radiation, and is intrinsic to the NaI detectors.  Given the expected count rate within a detector, it will take $\sim 2$ years to improve upon the measured half life for \ti{}.  However, we will be able to measure the shorter-lived isotopes, such as \co{} to $0.005\1{y}$ after only one year.

This experiment uses a total of thirty-two 76B76/3mL-E1 NaI(Tl) detectors from Scionix. These detectors are each composed of cylindrical $76\1{mm} \times 76\1{mm}$~NaI(Tl) scintillator crystals coupled to ETL9305 photomultiplier tubes (PMTs). Since NaI(Tl) is one of the most commonly used scintillators, the environmental influences derived from this experiment can be expected to be of broad relevance. These scintillators have a high light yield of $4$\scinot{4}$\1{photons/MeV}$~\cite{knoll2000} and the senstivity of the bialkali photocathodes are well matched to the scintillation wavelength. Each detector is equipped with an LED that is used to perform an \textit{in situ} deadtime measurement. The PMTs are biased using CAEN DT5533P high voltage power supplies. These power supplies enable continuous monitoring of the high voltage and current powering each detector. The operating high voltages are set such that the energy resolution is independent of high voltage near the operating point. 

\begin{figure}[!htbp]
\begin{center}
\includegraphics[width=0.55\columnwidth]{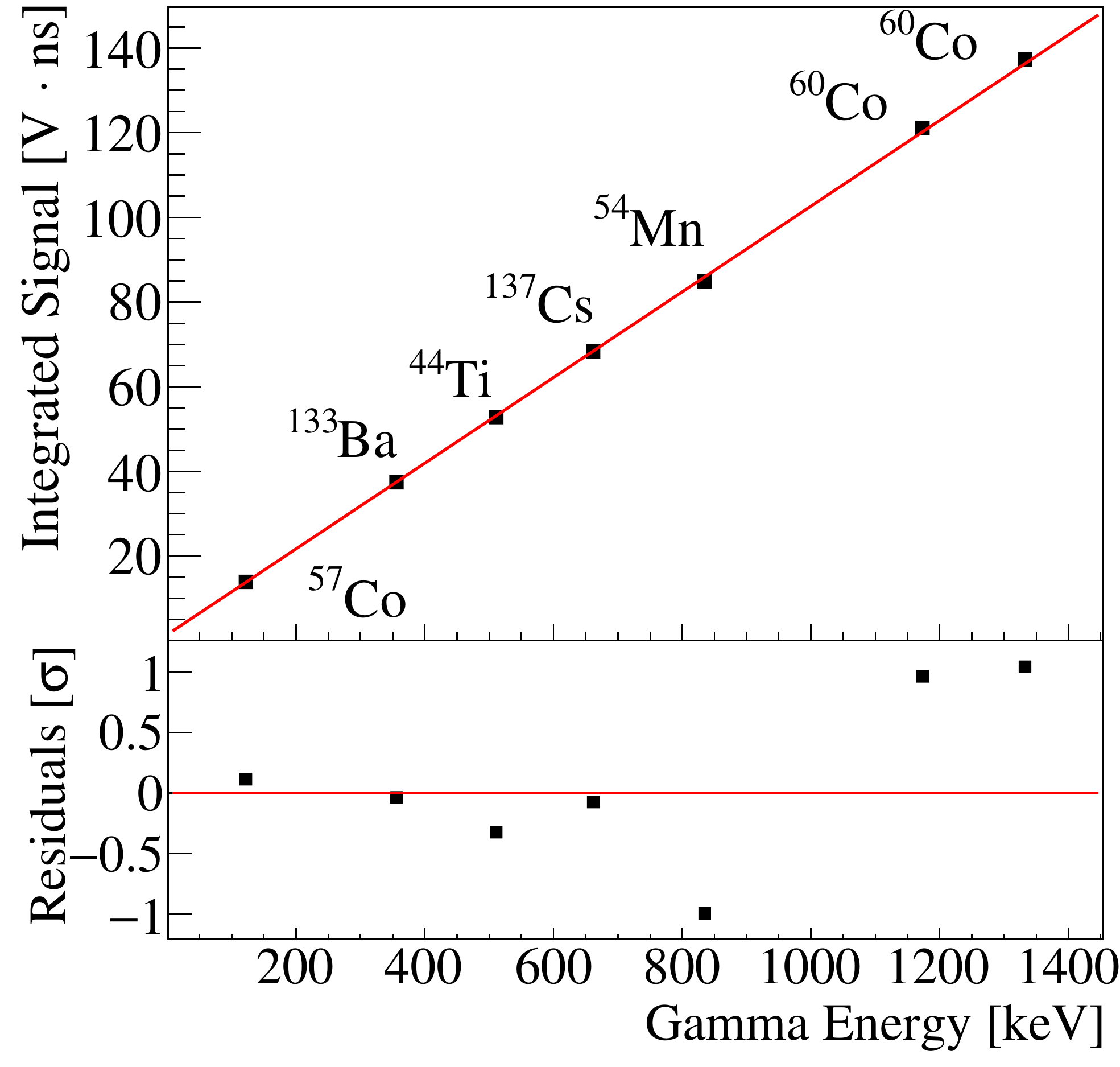}
\end{center}
\vspace{-5mm}
\caption{Energy calibration for one NaI(Tl) detector, using gamma reference sources. The relationship between energy and the mean of the photopeak is linear over the range of interest. Uncertainties are quoted as the width of the fitted Gaussian over the statistical error, which are smaller than the marker size shown. \label{energycalib}}
\end{figure}

To confirm the linear response of the detectors~\cite{Engelkemeir:1956}, we record energy spectra for all detectors from a variety of gamma sources and measure the position of the photopeak as the center of a fitted Gaussian distribution. The uncertainties are quoted as $\sigma / \sqrt{N}$ where $\sigma$ is the width of the fitted Gaussian in the fit range, and $N$ is the number of counts within that distribution. An example energy calibration for a representative detector is shown in Fig.~\ref{energycalib}. We find that all our detectors exhibit a linear response well beyond the gamma emission lines of our sources. In order to accurately account for possible variations, we perform an additional check of energy calibration using the available reference source every four hours. 

\begin{figure}[!htbp]
\begin{center}
\includegraphics[width=0.55\columnwidth]{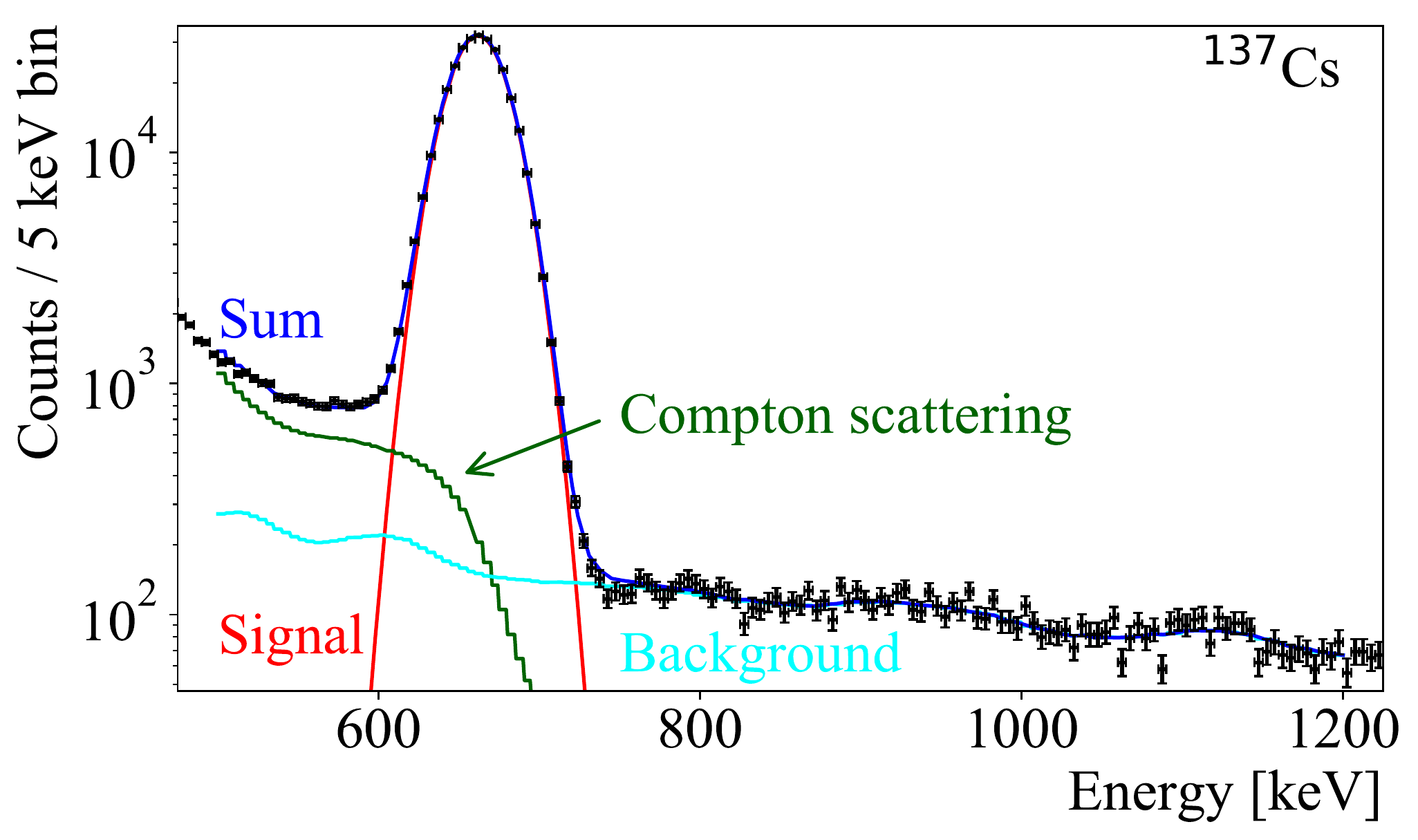}
\end{center}
\vspace{-5mm}
\caption{This example \cs{} spectrum was measured by a single detector with $3600\1{s}$ livetime.  A fitting routine is used to extract the rate inside of the full-absorption peak. The fit components include signal, Compton-induced scattering, modeled by Geant4 and the measured background simultaneously in order to measure the contributions of each of these components. 
\label{cs137resid}}
\end{figure}

In order to extract the decay rate of a source, we use a fitting routine to obtain the number of counts inside the full absorption peak as shown in Fig.~\ref{cs137resid}, which takes into account the background at each location (as measured by the dedicated detector pair), a Geant4 simulation of the Compton-induced background from the monitored source, and the photopeak. The rate is calculated by integrating the photopeak and dividing by the livetime. 

\begin{figure}[!htbp]
\begin{center}
\includegraphics[width=0.6\columnwidth]{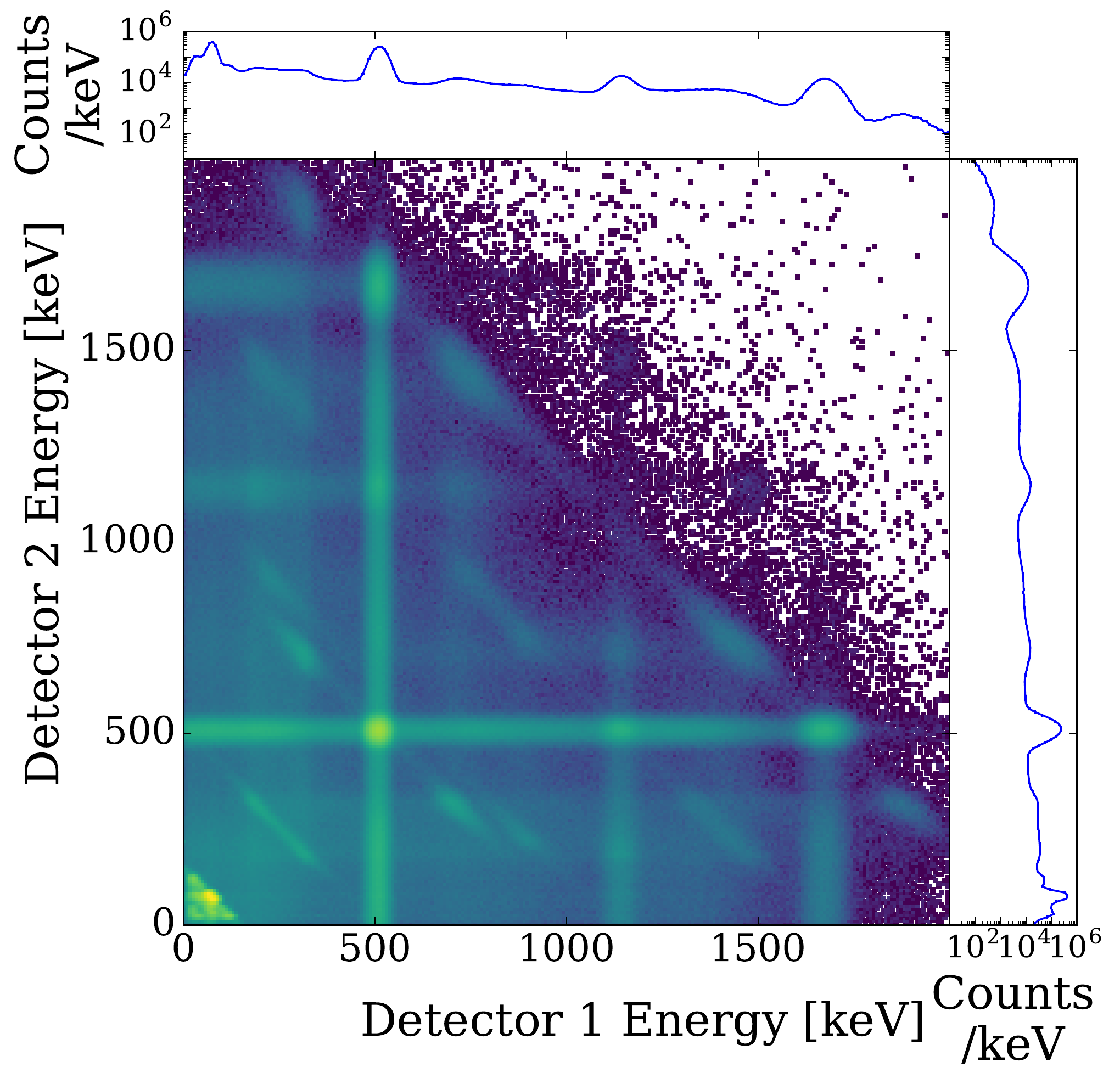}
\end{center}
\vspace{-5mm}
\caption{Coincident energy depositions seen in a detector pair monitoring the same \ti{} source. The coincidence requirement is a time offset $\leq 20\1{ns}$ of the trigger time in either detector. The $511\1{keV}$ pair production peak is visible, along with the 1157 keV full absorption peak, and the sum of those two peaks (1668 keV). Associated Compton scatterings are visible as diagonal lines intersecting the respective photopeak. Additionally, at low energies the X-rays at $68$ and $78\1{keV}$ from the daughter of \ti{} are also visible. 
The color scale is logarithmic and the spectra of the individual detectors are projected along their respective axes. \label{fig:ti44coin} }
\end{figure}

Our high angular acceptance, coupled with accurate timing (cf. Sec.~\ref{sec:DAQ}) allows for detailed coincidence studies. Analysis of Compton scatters show coincident events for two detectors, as demonstrated in Fig.~\ref{fig:ti44coin} with a \ti{} source. Various combinations of full absorption and Compton-based energy deposition can thus be studied. We plan to use these cross-detector correlations to examine geometry-related rate changes in future measurements. Further, we foresee a muon veto based on scintillation tiles coupled to silicon photomultipliers with dedicated on-board single-channel electronics integrated with the existing DAQ. The veto will enclose the NaI(Tl) detectors to monitor the cosmic ray muon flux, and will allow event-by-event identification of muon-induced events as opposed to radioisotope decays.

\section{Data Acquisition System and Data Handling} \label{sec:DAQ}

\subsection{Waveform Digitization} \label{sec:trigger}

One of the key features of our experimental design is that each event is individually stored to enable event-by-event correlation studies. To achieve this, the full voltage trace (waveform), as opposed to a time-binned trigger rate, is digitized and stored individually using a custom data acquisition system (DAQ). Among other correlation studies, we intend to perform cross-correlation studies between detectors, so the DAQ must be able to sample all channels simultaneously, while keeping deadtime to a minimum. With the DAQ presented in this section, we are able to sample all available detectors simultaneously in a deadtime-free manner.

We use an NI5751 Analog-to-Digital Converter (ADC), which samples up to 16~channels at a rate of 50\,MS/s per channel with 14\,bit resolution. Each channel has a voltage range of $\pm2\1{V}$ with a bandwidth of $26\1{MHz}$. An NI-PXI-7951R Field Programmable Gate Array (FPGA) temporarily stores the data from each channel into individual First In First Out (FIFO) circular buffers. Once a trigger is detected, the waveform trace is stored until it is streamed via fiber optic cable to a host computer, and written to disk. Every waveform is stored and available for offline inspection. 

\begin{figure}[!htbp]
\begin{center}
\includegraphics[width=0.55\columnwidth]{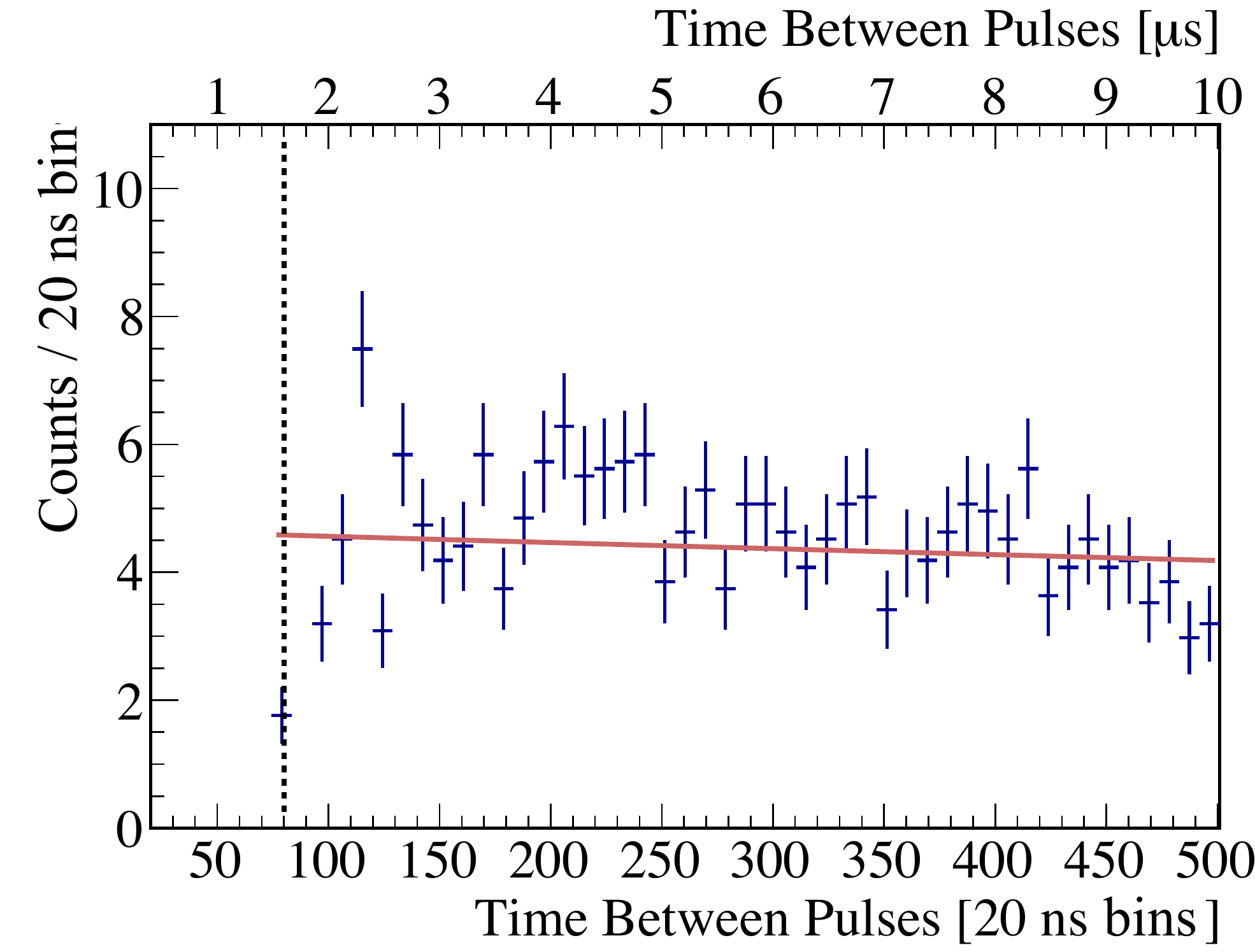}
\end{center}
\vspace{-5mm}
\caption{Time between consecutive events in $20\1{ns}$ bins. The minimum time between events is exactly the length of the post-trigger region, as indicated by the vertical line. The expectation for the time distribution is given without any free parameters as the red line and is consistent with the observed data. \label{fig:deadtime}}
\end{figure}

Voltage is read continuously into the DAQ channels, and is binned according to ADC clock cycles. In order to trigger the DAQ, a waveform bin must: (1) be greater than the bin before it (rising edge), (2) pass a user-specified threshold voltage and (3) not be within $\sim7$ times the NaI(Tl) decay time ($1.6\1{\mu s}$) in order to minimize the occurrence of pile up events. Thus, in order for a trigger to occur, the waveform bin cannot already have been recorded as part of another waveform. 
Because of our circular buffer, even when a trigger has been detected, the ADC is still acquiring data. Fig.~\ref{fig:deadtime} shows the time between consecutive events in a given channel. The minimum time between two consecutive events is the previously specified $7$ times the NaI(Tl) decay time, substantiating our claim that the system is deadtime free. 

\begin{figure}[!htbp]
\begin{center}
\includegraphics[width=0.55\columnwidth]{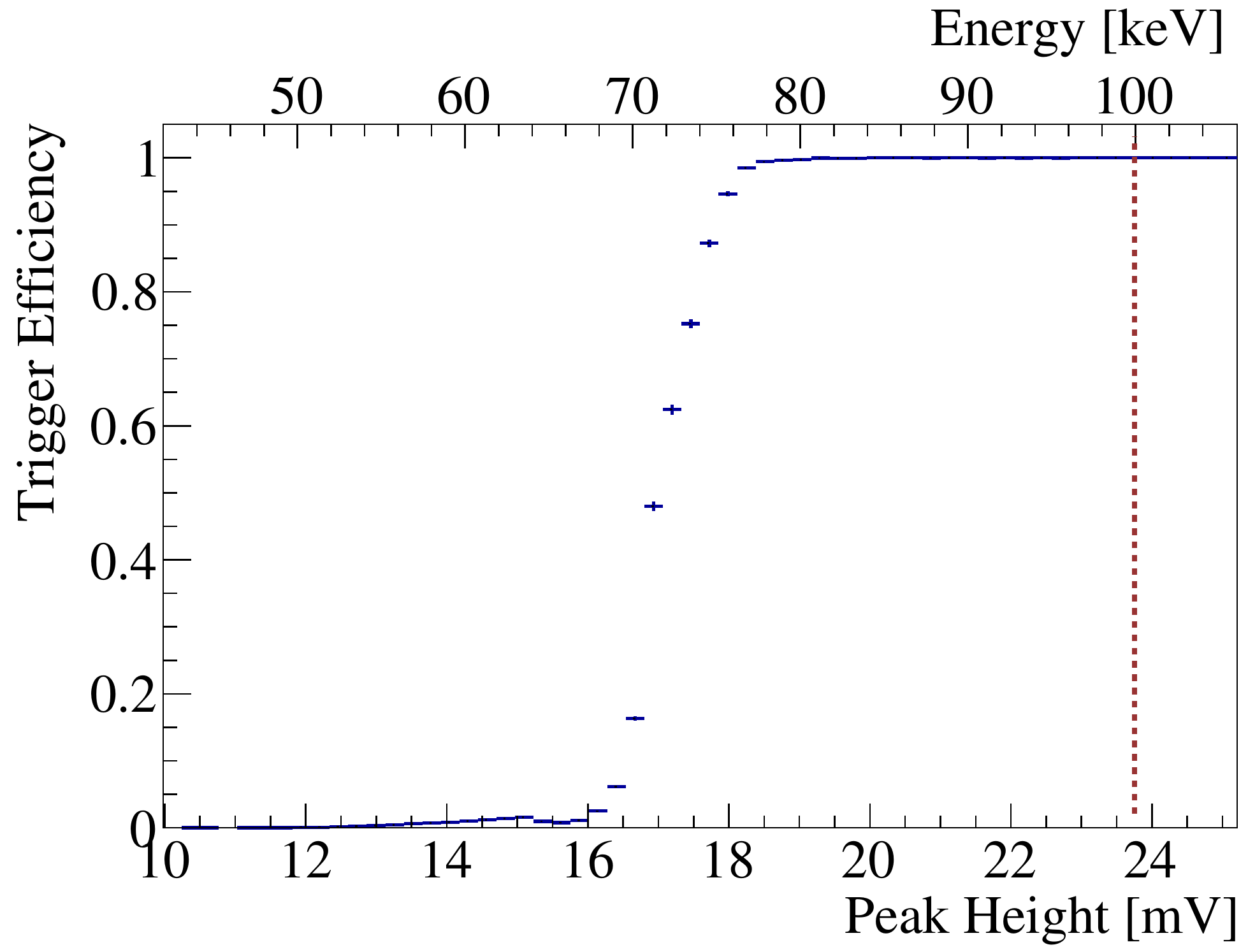}
\end{center}
\vspace{-5mm}
\caption{Trigger roll-off for a NaI(Tl) detector. The bottom axis corresponds to the peak height recorded by the ADC, while the top axis is the calibrated energy scale. The typical trigger threshold is indicated by the vertical line. \label{fig:rolloff}} 
\end{figure}

To measure the trigger efficiency, we temporarily configured the FPGA logic such that one detector in a pair acts as the external trigger for the other detector.  We then measure the fraction of events in the other detector which would have triggered as a function of the energy of these events, as shown in Fig.~\ref{fig:rolloff}.  At energies above $100\1{keV}$, we observe no missed triggers, which yields a total fraction of $<6$ \scinot{-7} missed triggers at a $99.7\%$ confidence level. To this end, we typically set our previously-mentioned threshold voltage to be $\sim 100\1{keV}$, which is well above the trigger rolloff shown in Fig.~\ref{fig:rolloff}. 

\subsection{LED Pulser}

\begin{figure}[!htbp]
\begin{center}
\includegraphics[width=0.55\columnwidth]{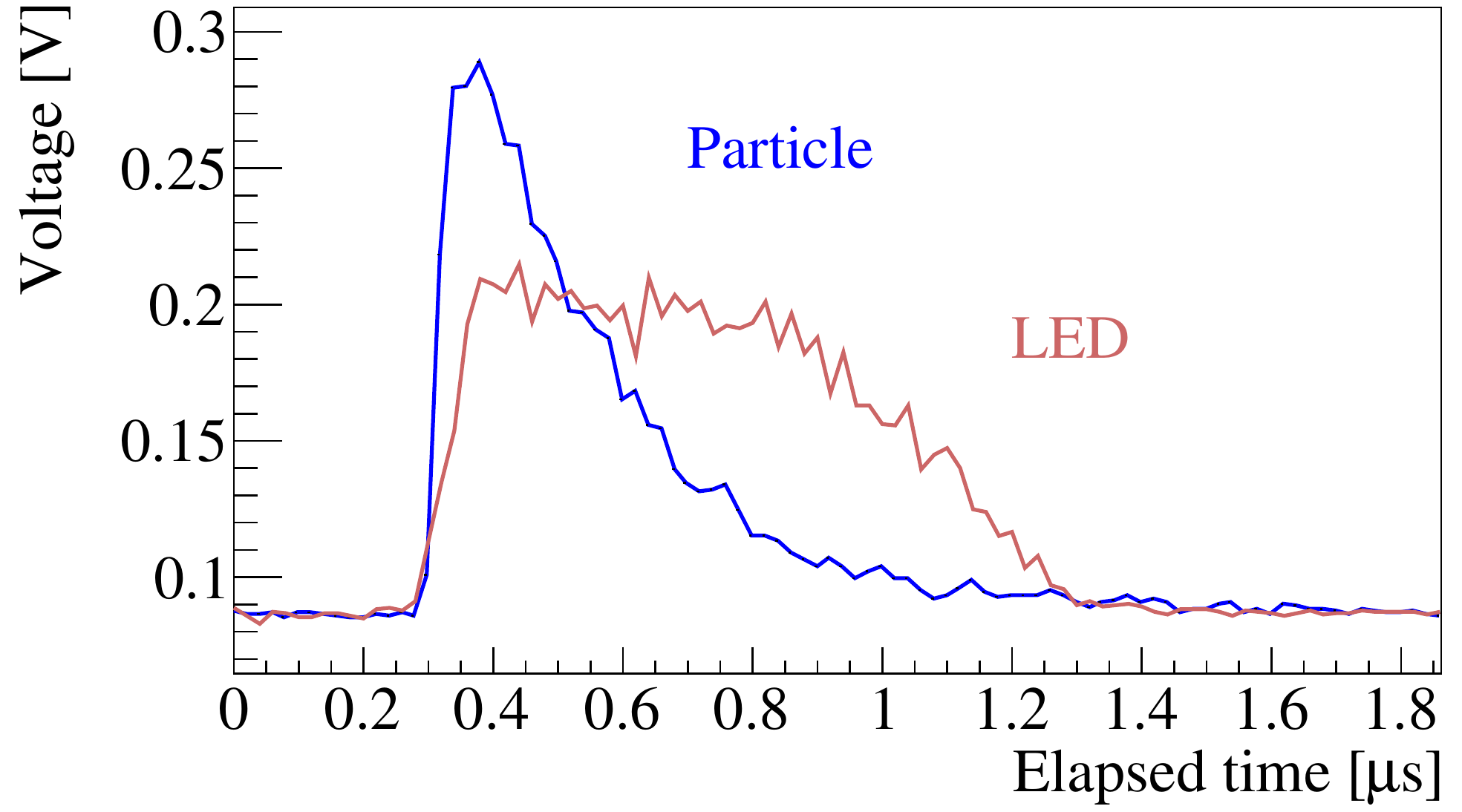}
\end{center}
\vspace{-5mm}
\caption{Waveform comparison between a particle pulse (an 834.8 keV $\gamma$ from $^{54}$Mn) and an LED pulse. The LED pulses are flagged by the DAQ for offline analysis. \label{fig:LEDwave}}
\end{figure}

\begin{figure}[!htbp]
\begin{center}
\includegraphics[width=0.6\columnwidth]{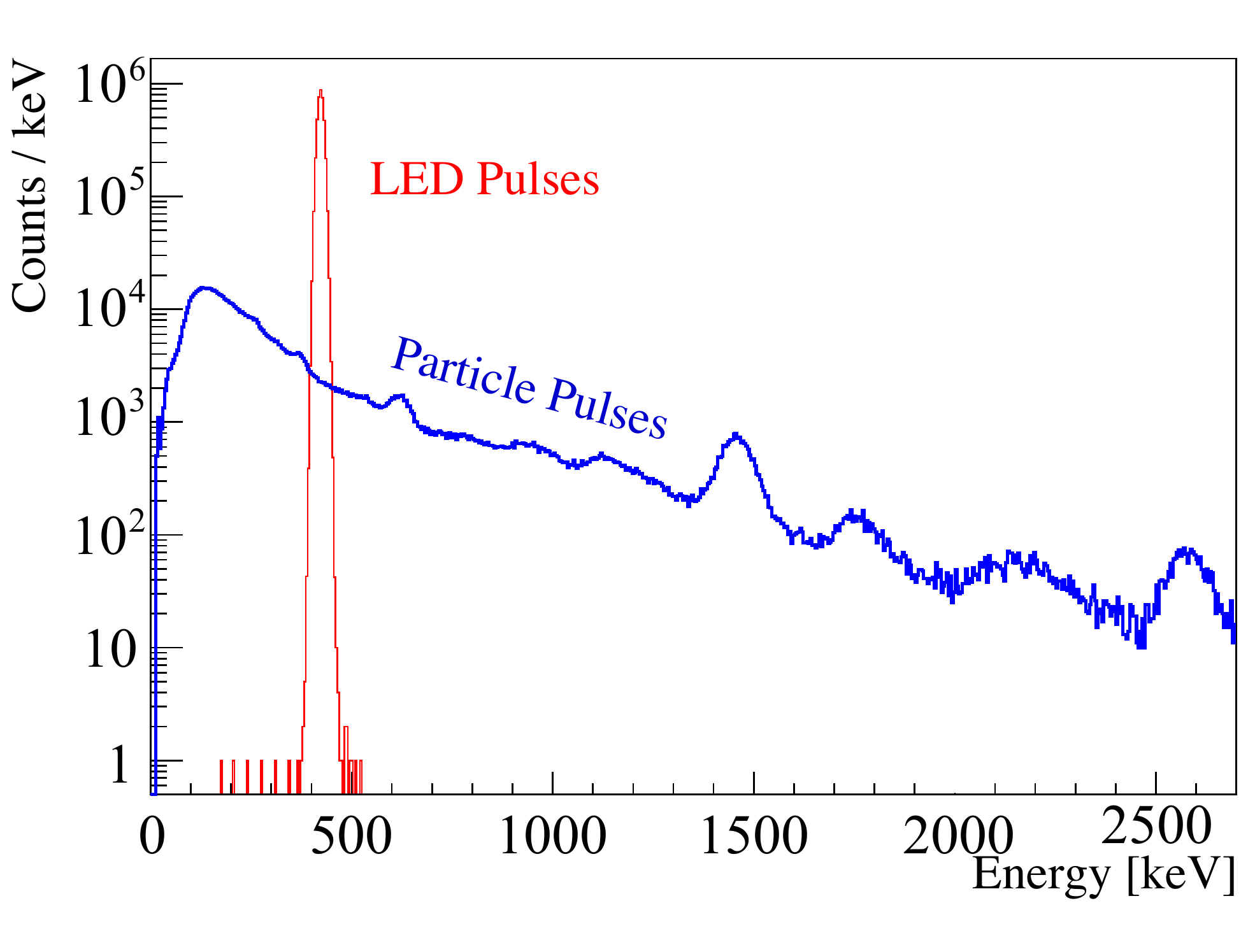}
\end{center}
\vspace{-5mm}
\caption{Background energy spectrum (particle pulses) compared to the LED spectrum flagged and separated. The LED pulses can be used to estimate the amount of deadtime in the setup and to crosscheck any analysis for time-dependent decay rates. \label{fig:ces_bg_led}}
\end{figure}

A $470\1{nm}$ LED pulser is used as an artificial reference standard in each detector to measure the efficiency of the entire setup, including the DAQ as well as the analysis chain. A square wave test pulse is sent from the DAQ through active analog circuitry, which is used to adjust the detected light level to create the response as shown in Fig.~\ref{fig:LEDwave}. Given the measured turnaround time between sending the pulse out and recovering it in the digitized data stream, the time at which the pulse occurred is flagged as part of the data stream, aligning with the detected LED waveform. These flags are embedded in the data, and can be used in later analyses. Fig.~\ref{fig:ces_bg_led} shows an energy spectrum of pulses flagged according to their origin as particle or LED pulses. Because we send the pulses to the LED directly, using the DAQ Labview code, we can compare the number of tagged pulses sent to the number of LED pulses observed at the analysis level. We find the probability that we do not tag an LED pulse to be $< 2$ \scinot{-8}, at a 99.7$\%$ confidence level. The LED is therefore a reliable reference source.

As discussed in Sec.~\ref{sec:trigger}, the minimum time between subsequent pulses is the artificially set $1.8\1{\mu s}$ event window. Since no events are lost to the DAQ digitizing events, we say the system is deadtime free. However, two events occurring in the same event window, referred to as pile up, is a related issue. Because the second event did not cause a trigger, this event typology could be considered a form of deadtime if not accounted for properly. Through use of LED tagging, we are able to characterize the nature of pile up, which can occur as a result of either two independent single scatter events, or as a result of a single multi-scatter event. Since we have a well-characterized population of LED events, any deviations in the expected energy indicate the presence of a pile up event at that energy range. We examine various event typologies to differentiate pile up events from single scatter events. As seen in Fig.~\ref{fig:LEDwave}, the LED waveform and a waveform from a particle pulse have slightly different shapes. Because of this, the LED was used primarily to identify event typologies and population distributions, rather than set cut definitions. Using cuts defined based on the linearity between the pulse height and integral of a waveform, the baseline of the pre-trigger region, and its RMS noise, we remove the pile up events, resulting in the spectrum shown in Fig.~\ref{fig:ces_bg_led}. Since the height of the LED can be changed, pile up at different energy ranges can be explored. In general, the LED peak is set to be inside the Compton continuum of a photopeak.

The LEDs are constantly pulsed at a rate of $\sim100\1{Hz}$, which is the typical rate at which we observe events from the absorption peaks of our sources. This results in a pile up rate of $1.6\pm0.5\1{Hz}$, which is removed from the population of single-scatter events at the analysis level. At the moment, the LED pulses are triggered at a constant rate, though development is underway for a Poisson-distributed rate, to better simulate the observed sources.

\subsection{Data Handling}
Each detector pair records events at a rate of about~$1.1\1{kHz}$, which includes $\sim100\1{Hz}$ in the full absorption peak and a similar rate from the LED pulser. Given that each waveform bin is stored as a 16~bit integer, every event is $200\1{bytes}$. Over the course of a year, we store $\sim44\1{TB}$ per setup. The raw slow control data (such as temperature or high voltage readings) is stored as well, but contributes a mere $4\1{GB}$ annually.

To process the data, we extract relevant information about each event including: the detector in which the event occurred, the Unix timestamp, the height and integral of the measured waveform, whether or not an LED is responsible for the event, and the most recent environmental readings. Processing the data yields a compression factor of $\sim 12$, once stored in ROOT trees~\cite{Brun1997}. 

\section{Slow Control and Slow Monitoring}

\subsection{Monitoring of Environmental Parameters}
\begin{figure}[!htbp]
\begin{center}
\includegraphics[width=0.65\columnwidth]{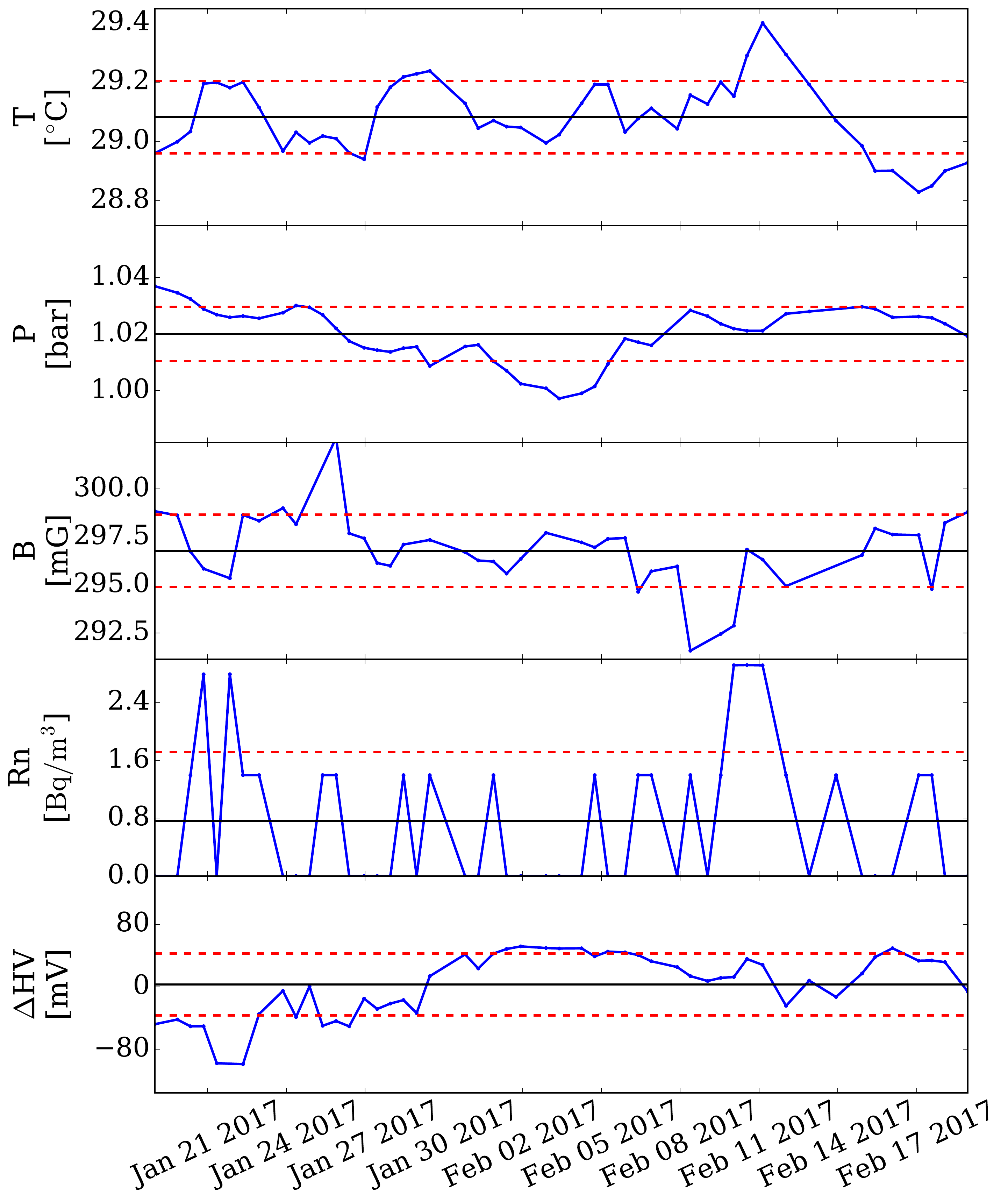}
\end{center}
\vspace{-5mm}
\caption{Example month of environmental and operational parameters, including: temperature (T), pressure (P), total magnetic field (B), radon level (Rn), and high voltage deviations from the set value in a representative channel ($\Delta$HV). The mean of each parameter is represented by a solid horizontal line, and RMS in dashed lines. Data was averaged in day-long bins. \label{fig:slowstab}}
\end{figure}

Temperature, pressure, humidity and magnetic field are measured with I2C sensors, which are monitored through an Arduino Uno board. This board is directly connected to the DAQ host machine via USB. High voltage levels are read through the CAEN high voltage supplies. Finally, the radon levels are monitored with a radon monitor, equipped with a Drystik unit for continuous operation. To ensure accurate timing, each of these measurements is Unix timestamped, and each host computer is synchronized with the NIST time server every $5\1{minutes}$. Thus, every event detected by a NaI(Tl) detector has a detailed account of the environmental conditions when it occurred. Fig.~\ref{fig:slowstab} shows an example month of data, taken with the temperature controlled and nitrogen flushing on. Tab.~\ref{table:arduino} details the average values for this month of data, along with the manufacturer details, RMS and the resolution of each sensor. 

\begin{table} \footnotesize
\begin{center}
\begin{tabular}{ lrrrrrr }
\hline
Parameter & Manufacturer & Model & Average & ~RMS & ~Resolution& Unit\\
\hline 
Temperature & Bosch & BMP085 & $29.2$ & 0.1 & $0.1$ &$^\circ$C \\
Pressure & Bosch & BMP085 & 1.02& 0.01& 3 \scinot{-5} &bar\\
Total Magnetic Field & Honeywell & HMC5883L & 301 & 12 & 12& mG \\
Relative Humidity & Hope RF & HH10D & $0$ &$0.7$ & $0.1$&$\%$ \\
High Voltage & CAEN & DT5533P & 600-1000 & 0.02 & 0.1 &V \\
Radon Activity & Durridge & RAD7 & $0.5$ & $1.1$ & $0.8$&~\radunits \\
\end{tabular}
\end{center}
\caption{Environmental and operational sensor details, including the manufacturer information and the parameters monitored by a setup at a rate of 1 $\rm{min}^{-1}$.}
\label{table:arduino}
\end{table}

\subsection{Temperature Control} \label{sec:tempcontrol}
\begin{figure}[!htbp]
\begin{center}
\includegraphics[width=0.55\columnwidth]{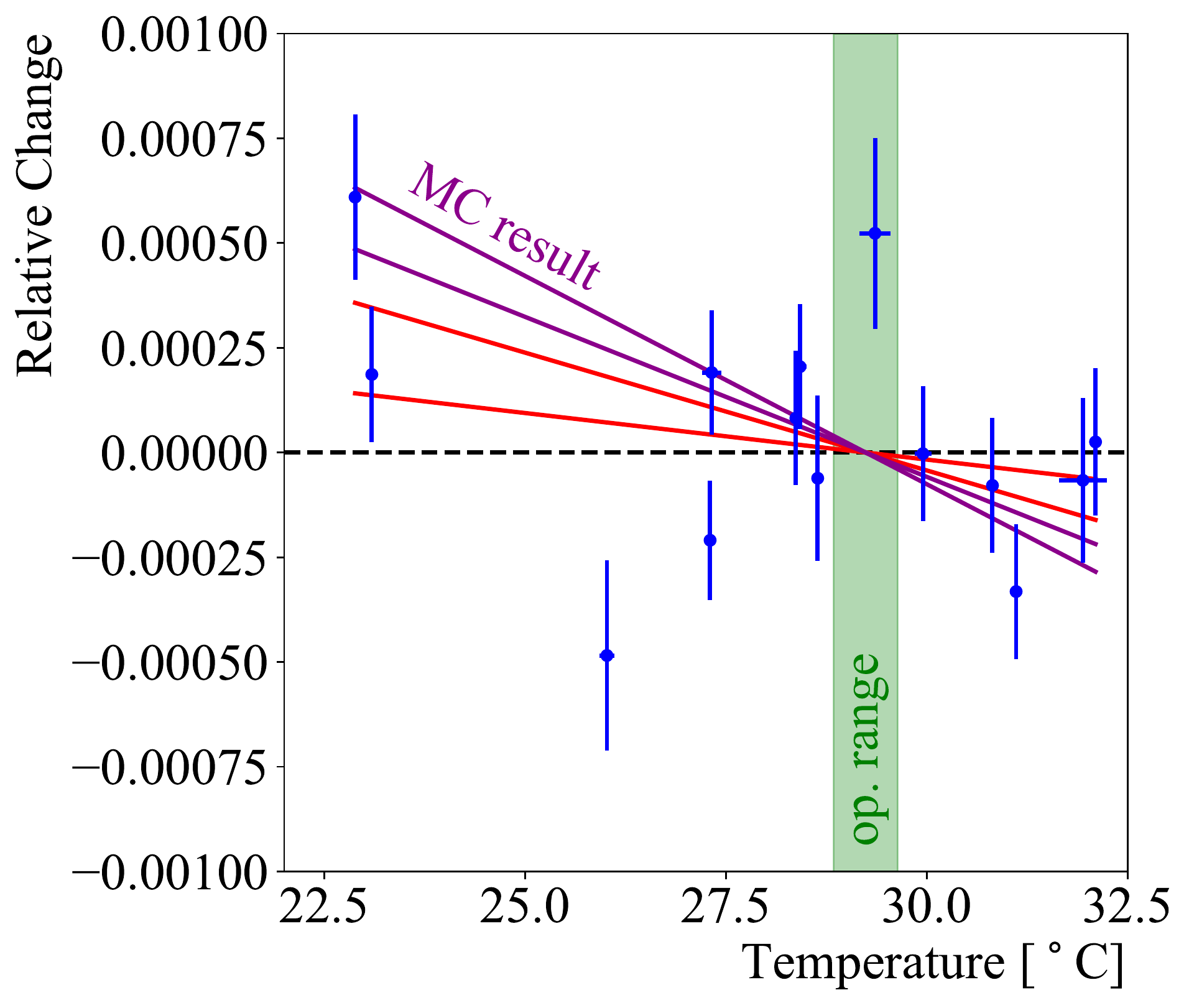}
\end{center}
\vspace{-5mm}
\caption{The change of the measured rate of \ti{} is shown with the $1 \sigma$ variation in the fitted slope (solid red lines), which represent the variation in rate expected from temperature variations for a setup.  The simulated change in the detection probability due to the thermal expansion of the aluminum frame shown in Fig.~\ref{fig:detectorset}, and its expected variation, is shown in purple. The default temperature region is based on Fig.~\ref{fig:slowstab}. \label{fig:ratevtempMC}} 
\end{figure}

\begin{figure}[!htbp]
\begin{center}
\includegraphics[width=0.4\columnwidth]{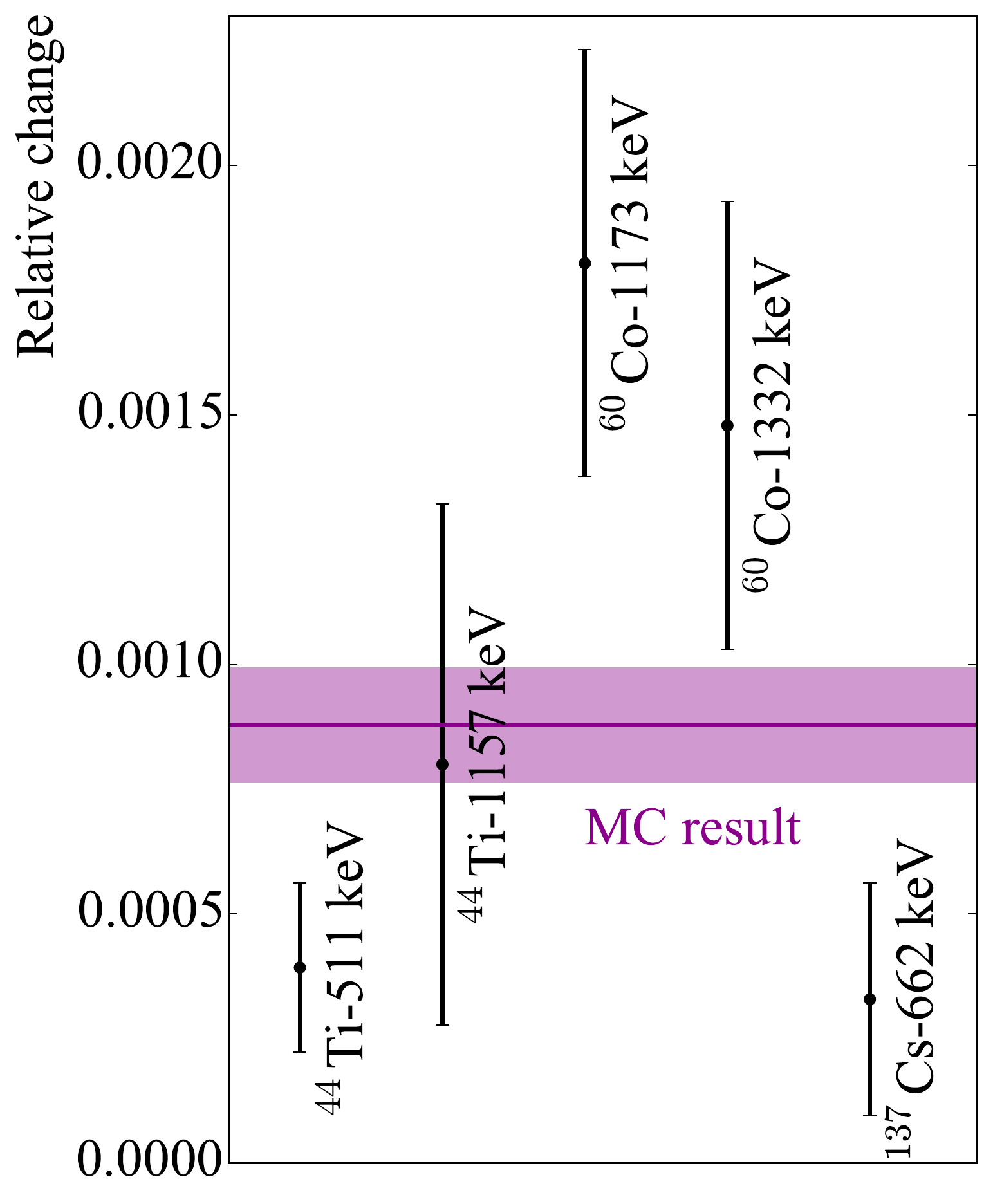}
\end{center}
\vspace{-5mm}
\caption{The effect of a deliberate $10\1{K}$ temperature decrease on the measured rate. Shown are the expectation from the thermal expansion simulation ($\pm1\sigma$) as well as measurements from several sources. 
\label{fig:MCresult}}
\end{figure}

The thermal expansion of the aluminum frame shown in Fig.~\ref{fig:detectorset} may have a systematic effect on the observed rate. An increase in temperature can cause the distance between the source and the detectors to change, which also changes the solid angle subtended by the NaI(Tl) crystals. We use a Monte Carlo model to evaluate the expected detection probability at different temperatures, by using an aluminum thermal expansion coefficient of $2.36 \times 10^{-5}\1{K^{-1}}$~\cite{Agilent:2012}. For a temperature change of \cel{10} the model predicts a relative rate change of $(8.8\pm1.2)$\scinot{-4}. The model is compared to source measurements as shown in Fig.~\ref{fig:ratevtempMC} where the temperature was changed deliberately to quantify the effect on the measured rate. Fig.~\ref{fig:MCresult} shows the change in measured rate for various sources given a \cel{10} temperature difference as in Fig.~\ref{fig:ratevtempMC}, compared to the expectation from this simulation. Though the Monte Carlo result and the fit from the measured rate do not agree within the stated $1\sigma$ statistical uncertainties, they do show the same overall trend. The thermal expansion of the aluminum frame should be considered when evaluating any correlations between temperature and rate fluctuations. 

In order to stabilize the temperature, the insulated inner box contains two PID-controlled heating pads ($76\times508$ mm) operating at $230\1{V}$ each, supplying $250\1{W}$ of power. Temperature monitoring is achieved through an I2C sensor located inside the inner box, with a resolution of \cel{0.1}. A safety system will cut off the power if the maximum allowed heater temperature is exceeded, to protect the detectors and electronics. A COMSOL simulation of the inner box, detector pairs, and lead bricks was performed in order to study the thermal distribution of the inner box. The temperature across the eight detectors does not vary more than \cel{1}, in agreement with measurements. As can be seen in Fig.~\ref{fig:slowstab}, the temperature is stable to within \cel{0.1}. 
 
\section{Limits on Systematic Influences}

\begin{figure}[!htbp]
\captionsetup[sub]{font = large, labelfont={bf, sf}}
\begin{subfigure}{0.45\textwidth}
\begin{center}
\includegraphics[width=\columnwidth]{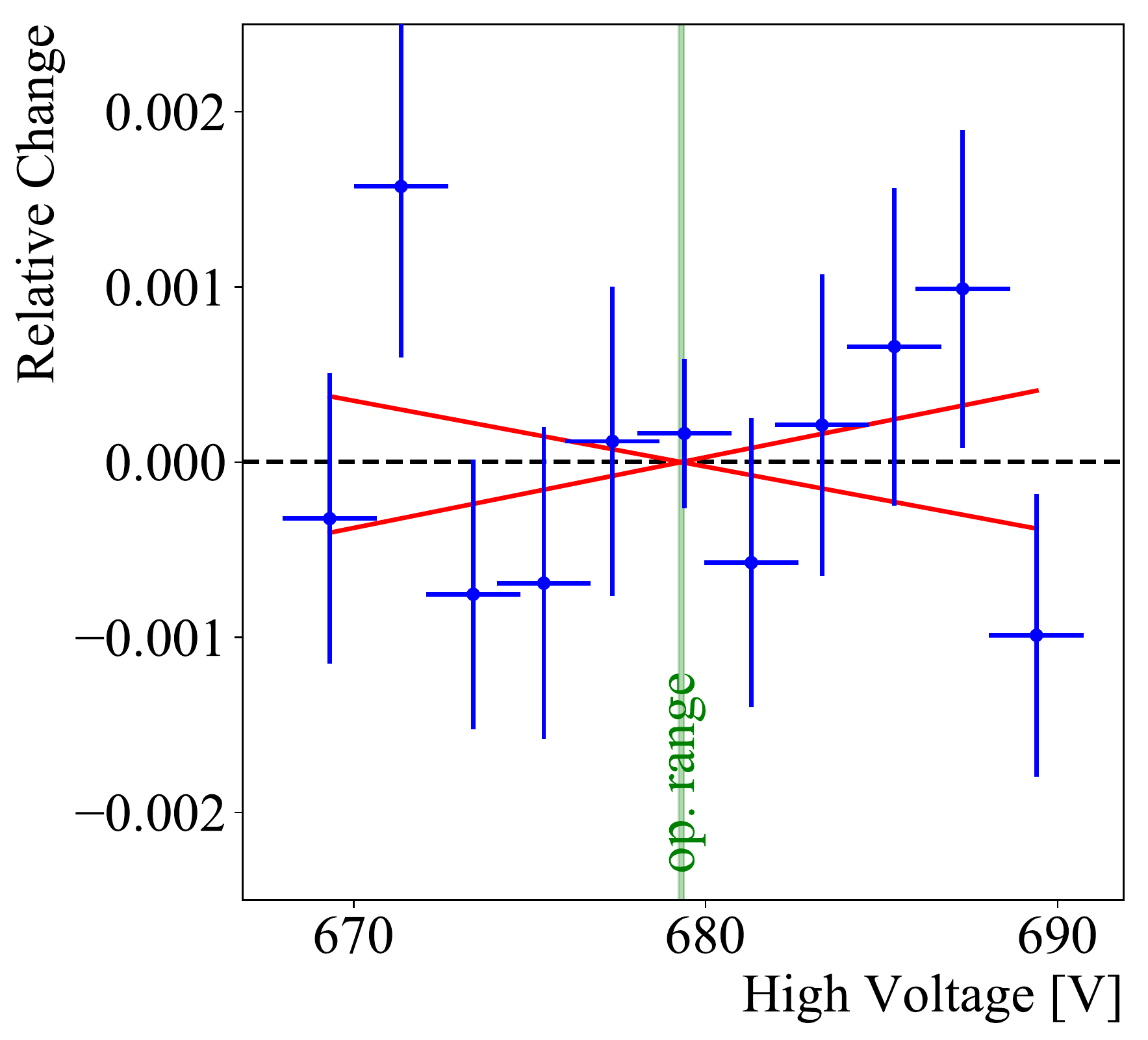}
\end{center}
\vspace{-0.92\textwidth}
\caption{\label{fig:ratevhv}}
\vspace{0.8\textwidth}
\end{subfigure}
\begin{subfigure}{0.45\textwidth}
\begin{center}
\includegraphics[width=\columnwidth]{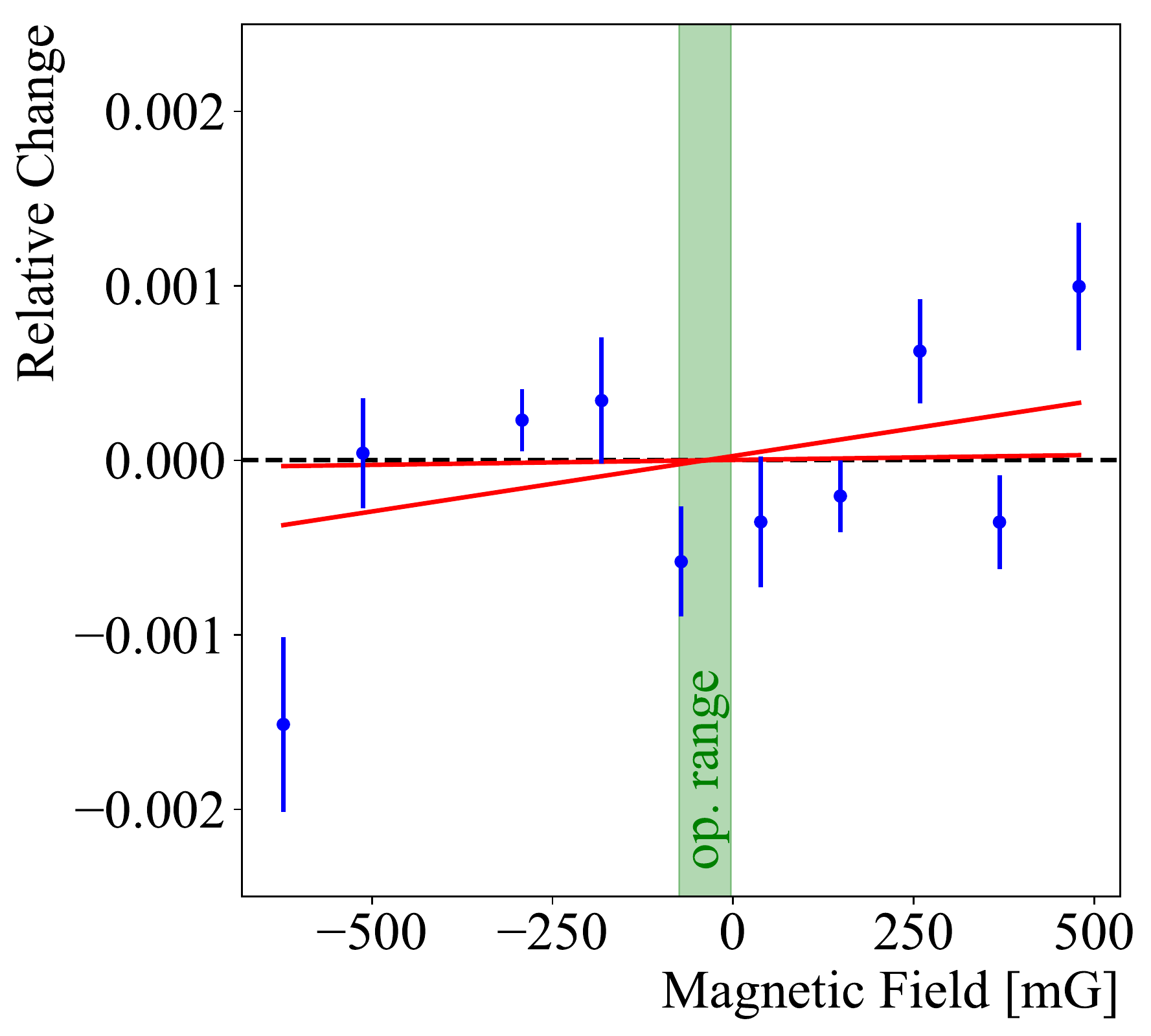}
\end{center}
\vspace{-0.92\textwidth}
\caption{\label{fig:ratevb}}
\vspace{0.8\textwidth}
\end{subfigure}
\begin{subfigure}{\textwidth}
\begin{center}
\includegraphics[width=0.45\columnwidth]{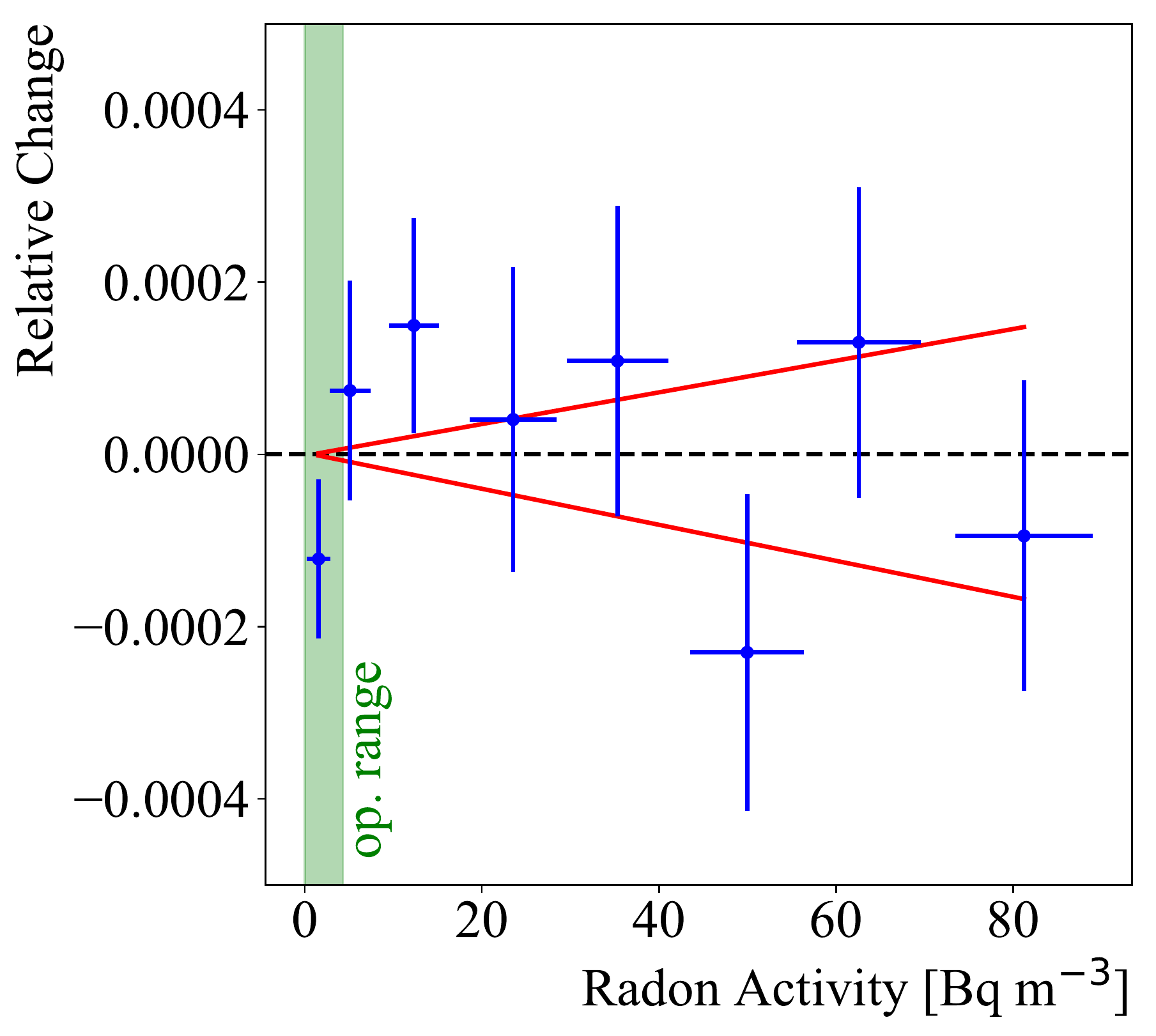}
\end{center}
\vspace{-0.42\textwidth}
\caption{\label{fig:ratevrad}}
\vspace{0.35\textwidth}
\end{subfigure}

\caption{Measured rate while deliberately varying magnetic field, ambient radon level and high voltage. Data was measured once while increasing the given parameter and again while decreasing it. While no statistically significant effect is visible in any of the measured parameters, the $1 \sigma$ variation in the fitted slope (solid red lines) allows for a combined variation of no more than \order{-4}, satisfying our design criteria. Typical operating ranges are shown as the vertical (green) shaded region. \label{fig:ratevslow}}
\end{figure}

In order to characterize the potential impact of systematic sources of error on the measured decay rate, we deliberately varied a given environmental parameter over a range much larger than its usual operating range. Interpolation of these measurements over the operating range places a limit on the maximum systematic uncertainty that can be attributed to this parameter. Tests were performed by individually scanning high voltage, temperature, magnetic field and radon concentration both up and down to detect potential hysteresis effects. Tab.~\ref{table:ratevslow} summarizes the results of these measurements as discussed below. For all of the parameters, no correlation between rate and the given environmental parameter is apparent, as seen in Figs.~\ref{fig:ratevtempMC} and \ref{fig:ratevslow}. 

While Figs.~\ref{fig:ratevslow} do not exhibit an obvious correlation between rate and an external parameter, there is also no physically-motivated model that predicts such a dependency. A Kolomogorov-Smirnov (K-S) test was performed on each measurement to test whether or not a given measurement is drawn from the expected Poisson distribution, taking into account the livetime of each measurement. In all cases the K-S test favors the null hypothesis, that each measurement is likely drawn from a Poisson distribution, indicating that there is no evidence for a rate-dependent effect. In the absence of any models of external influences on the measured decay rate, we fit a linear function as suggested in~\cite{Towers2013} and infer the largest possible variation that can be seen over the expected operating range. The expected operating range of each environmental parameter is taken to be three times its RMS value (taken from Tab~\ref{table:arduino}), as seen in Fig.~\ref{fig:slowstab}. 

Fig.~\ref{fig:ratevtempMC} shows the change in measured rate of a source when the temperature is varied. Given the simulation result and that the setup is capable of stabilizing the temperature to within \cel{0.4} over extended periods of time (see Sec.~\ref{sec:tempcontrol}),
the expected systematic effect is $(1.7 \pm 0.7$) \scinot{-5} for the considered temperature variation (Tab.~\ref{table:ratevslow}). The K-S test indicates that this measurement is consistent with a Poisson distribution, with a p-value of 0.96. Though the thermal expansion of the aluminum frame is a systematic effect, the actual measurement of the rate as a function of temperature indicates we are not sensitive enough for this effect to be dominant.

Fig.~\ref{fig:ratevhv} shows apparent rate variations as function of applied high voltage. No variation as function of high voltage is observed. While the PMT gain is a function of the applied high voltage, the spectral fit properly takes this into account. Thus, we find a limit on the maximally possible systematic variation of the measured fractional decay rate as function of high voltage to be 
$<2.2$\scinot{-6} at $99.7\%$ confidence level given the measured stability of the high voltage supplies of $0.07\1{V}$. Indeed, the possibility that this measurement is drawn from a Poisson distribution is favored, with a p-value of 0.93. 

In order to measure the effect of magnetic field variations, a Helmholtz coil was placed partially inside the inner box, enabling a range of $-600$ to $500\1{mG}$ to be applied to a detector pair. Because the height of the Helmholtz coil did not allow the inner box to close, these measurements were performed in an air-conditioned laboratory conditions. No statistically significant temperature or pressure excursions were detected during these measurements. A variation of the magnetic field over more than the operating range yields the data shown in Fig.~\ref{fig:ratevb}. The likelihood that this measurement originates from a Poisson distribution is favored, with a p-value of 0.98. A linear fit limits the possible systematic uncertainty from this environmental parameter to $(1 \pm 1)$ \scinot{-5} for an operating range of of $36\1{mG}$. 

Given the low radon concentration that results from flushing the inner boxes with nitrogen gas, our radon monitors only provide upper limits on the radon concentration and thus serve as a control of the integrity of the setups. While an increase in ambient radon concentration does result in an increased trigger rate, the spectral fit is not affected by a radon-increased background. We therefore limit the uncertainty from radon to $ <6.2$ \scinot{-6} at $99.7\%$ C.L.(Fig.~\ref{fig:ratevrad}) given the operating range of $<3.4$~\radunits. This measurement is again likely to be consistent with a Poisson distribution, with a p-value of 0.83. 

Due to the difficulty of enclosing a full detector pair inside a pressure chamber, it was not possible to measure the pressure dependence of the rate directly. However, given the $10\1{mbar}$ pressure variation observed of data acquisition shown in Fig.~\ref{fig:slowstab}, and since the other variables do not exhibit any rate dependent effects, we can estimate the maximum influence of the pressure using this dataset.  We find that the maximum possible variation due to pressure is $(0.7 \pm 1.1)$\scinot{-6}. While the relative humidity is continuously monitored, the standard operation condition with nitrogen-purged boxes leads to values below the sensitive range of the sensor, which such mainly serves to check the integrity of the setup. Overall, we find that possible systematic rate variations induced by environmental parameters are under control at the level of <\order{-5} in our experiment. The results of the dedicated measurements to study their influence are summarized in Tab.~\ref{table:ratevslow}.

\begin{table*} \footnotesize
\begin{center}
\begin{tabular}{ l l r r l l l r l r l }
\hline
Parameter & K-S     & Interpolated & Operating            & Maximum Systematic \\ 
          & p-value & Dependence   & Range ($3\times$RMS) & Influence \\
\hline 
Temperature 	& 0.96 & $(-4 \pm 2)$\scinot{-5}$\1{K}^{-1}$ 	& $ 0.4^\circ\1{C}$ &	$(1.7\pm0.7)$\scinot{-5}\\
Magnetic field 	& 0.98 & $(4 \pm 3)$\scinot{-7}$\1{mG}^{-1}$ 		& $36\1{mG}$ 	    & 	$(1 \pm 1)$\scinot{-5} \\
High voltage	& 0.93 & $(0 \pm 4)$\scinot{-5}$\1{V}^{-1}$ 			& $0.07\1{V}$ 	    & 	$< 2.2$\scinot{-6} at $99.7\%$ C.L. \\
Radon activity	& 0.83 & $(-0.1 \pm 2)$\scinot{-6}$\1{(Bq/m^3)}^{-1}$  	& $<3.4$~\radunits   & 	$< 6.2$\scinot{-6} at $99.7\%$ C.L. \\

\end{tabular}
\end{center}
\caption{Maximum systematic influence on the measured rate as a function of environmental and operational parameters. The rate was measured while varying temperature, 
magnetic field, radon level and PMT high voltage, as shown in Figs.~\ref{fig:ratevtempMC} and~\ref{fig:ratevslow}. The p-values of a Kolomorgov-Smirnov (K-S) test show that each measured parameter is favored to be consistent with a Poisson distribution without external rate-dependent influences. To determine the maximum possible influence on the measured decay rate, data was fitted with a linear relationship, giving the interpolated dependence shown above. The operating range of these parameters is shown in Fig.~\ref{fig:slowstab} and they are symmetric about a measured value. The maximum possible systematic influence on the measured decay rate is indicated.
\label{table:ratevslow}}
\end{table*}

\section{Conclusions and Outlook}
We have presented the design and characterization of an experiment built to carefully investigate the decays of long-lived radioactive isotopes. The design features control of environmental and operational parameters, allowing us to directly measure how changes in a parameter may affect the measured decay rates. We show that measured decay rates are not affected by systematics at levels $<\mathcal{O}(10^{-5})$ which is well below the reported anomalies of \order{-3}~\cite{Jenkins2009}. We will perform long term, stable measurements of the decay rates of various isotopes. This will deliver precise half life measurements and will allow us to test existing claims of variable decay rates. With four setups at different geographic locations, we can search for seasonal correlations, location-specific effects, and any cross-correlations. Fully digitized waveforms are available to permit extended investigation of the nature of events. By recording the Unix timestamps of individual events, we are able to calculate the power spectrum of each decay with periods spanning sub-seconds all the way to the duration of the measurements, expected to be multiple years. 

\section{Acknowledgments}
We are grateful for the support of the Purdue Research Foundation, the Netherlands Organization for Scientific Research, the University of Zurich, the Swiss National Science Foundation under Grants Nos. 200020-162501 and 200020-175863, the European Unions Horizon 2020 research and innovation programme under the Marie Sklodowska-Curie grant agreements No 690575 and No 674896, the Albert Einstein Center for Fundamental Physics at the University of Bern, UniBern Forschungsstiftung, FAPERJ and CNPq. We are very grateful to Rob Walet for designing the setup and to Rugard Dressler, Dorothea Schumann and Tanja Stowasser (PSI, Villigen, Switzerland) for preparing the \ti{} sources.

\bibliographystyle{JHEP}

\begin{thebibliography}{10}

\bibitem{Rutherford1910}
E.~Rutherford and H.~Geiger, \emph{\textit{LXXVI.} {The} probability variations
  in the distribution of $\alpha$ particles}, {\emph{Philosophical Magazine
  Series 6} {\bfseries 20} (1910) 698} [\href{https://doi.org/10.1080/14786441008636955}{\ttfamily DOI: 10.1080/14786441008636955]}.

\bibitem{meyer1927}
S.~Meyer and E.~Schweidler, \emph{Radioaktivit{\"a}t}. B.G. Teubner, Leibzig.,
  2~ed., 1927.

\bibitem{kohlrausch1928}
K.~W.~F. Kohlrausch, \emph{Radioaktivit{\"a}t Handbuch der Experimentalphysik},
  vol.~15. Akad. Verlagsgesellschaft mbH, Leibzig., 1928.

\bibitem{bothe1933}
W.~Bothe, \emph{Handbuch der Physik}, vol.~22 of \emph{1}. Springer-Verlag
  Berlin Heidelberg, Leibzig., 1~ed., 1933,
  [\href{https://doi.org/10.1007/978-3-642-90777-7}{\ttfamily DOI: 10.1007/978-3-642-90777-7}].

\bibitem{Goodwin:2009}
J.~R. Goodwin, V.~V. Golovko, V.~E. Iacob and J.~C. Hardy, \emph{Half-life of
  the electron-capture decay of $^{97}\mathrm{Ru}$: Precision measurement shows
  no temperature dependence}, {\emph{Phys. Rev. C} {\bfseries 80} (2009) 045501}  [\href{https://doi.org/10.1103/PhysRevC.80.045501}{\ttfamily DOI: 10.1103/PhysRevC.80.045501}].

\bibitem{Emery:1972}
G.~T. Emery, \emph{Perturbation of nuclear decay rates}, {\emph{Annual Rev.
  of Nucl. Sci.} {\bfseries 22} (1972) 165} [\href{https://doi.org/10.1146/annurev.ns.22.120172.001121}{\ttfamily DOI: 10.1146/annurev.ns.22.120172.001121}].

\bibitem{Hopke:1974}
P.~K. Hopke, \emph{Extranuclear effects on nuclear decay rates}, {\emph{Journal of Chemical Education} {\bfseries 51} (1974) 517}
  [\href{http://dx.doi.org/10.1021/ed051p517}{{\ttfamily DOI: 10.1021/ed051p517}}].

\bibitem{Hahn:1976}
H.~Hahn, H.~Bobn and J.~Kim, \emph{Survey on the rate perturbation of nuclear
  decay}, {\emph{Radiochimica Acta.} {\bfseries 23} (1976) 23} [\href{https://doi.org/10.1524/ract.1976.23.1.23}{\ttfamily DOI: 10.1524/ract.1976.23.1.23}].

\bibitem{Dostal:1977}
K.~P. Dostal, M.~Nagel and D.~Pabst, \emph{Variations in nuclear decay rates},
  {\emph{Zeitschrift f{\"u}r
  Naturforschung A} {\bfseries 32} (1977) 345} [\href{https://doi.org/10.1515/zna-1977-3-426}{\ttfamily DOI: 10.1515/zna-1977-3-426}].

\bibitem{Liu:2000}
L.~Liu and C.~Huh, \emph{Effect of pressure on the decay rate of
  $^{7}\mathrm{Be}$},{\emph{Earth and Planetary Science Letters} {\bfseries 180} (2000) 163 } [\href{http://dx.doi.org/10.1016/S0012-821X(00)00153-9}{\ttfamily DOI: 10.1016/S0012-821X(00)00153-9}].

\bibitem{Norman:2001}
E.~Norman, G.~Rech, E.~Browne, R.-M. Larimer, M.~Dragowsky, Y.~Chan et~al.,
  \emph{Influence of physical and chemical environments on the decay rates of
  $^{7}\mathrm{Be}$ and $^{40}\mathrm{K}$}, {\emph{Physics
  Letters B} {\bfseries 519} (2001) 15 } [\href{https://doi.org/http://dx.doi.org/10.1016/S0370-2693(01)01097-8}{\ttfamily DOI: 10.1016/S0370-2693(01)01097-8}].

\bibitem{Ellis1990}
K.~J. Ellis, \emph{The effective half-life of a broad beam {$^{238}$Pu/Be}
  total body neutron irradiator}, {\emph{Physics in Medicine and Biology}
  {\bfseries 35} (1990) 1079}.

\bibitem{Shnoll1998}
S.~E. Shnoll, V.~A. Kolombet, E.~V. Pozharskii, T.~A. Zenchenko, I.~M. Zvereva
  and A.~A. Konradov, \emph{Realization of discrete states during fluctuations
  in macroscopic processes}, {\emph{Physics-Uspekhi} {\bfseries 41} (1998)
  1025}.

\bibitem{Baurov2007}
Y.~A. Baurov, Y.~G. Sobolev, Y.~V. Ryabov and V.~F. Kushniruk,
  \emph{Experimental investigations of changes in the rate of beta decay of
  radioactive elements}, {\emph{Physics of Atomic
  Nuclei} {\bfseries 70} (2007) 1825} [\href{https://doi.org/10.1134/S1063778807110014}{\ttfamily DOI: 10.1134/S1063778807110014}].

\bibitem{Parkhomov2010A}
A.~G. {Parkhomov}, \emph{{Researches of alpha and beta radioactivity at
  long-term observations}}, {\emph{ArXiv e-prints} (2010) }
  [\href{https://arxiv.org/abs/1004.1761}{{\ttfamily arXiv: 1004.1761}}].

\bibitem{Parkhomov2010B}
A.~{Parkhomov}, \emph{{Periods Detected During Analysis of Radioactivity
  Measurements Data}}, {\emph{ArXiv e-prints} (2010) }
  [\href{https://arxiv.org/abs/1012.4174}{{\ttfamily arXiv: 1012.4174}}].

\bibitem{Baurov:2013}
Y.~A. {Baurov}, A.~Y. {Baurov}, A.~Y. {Baurov}, V.~A. {Nikitin}, V.~B. {Dunin},
  V.~V. {Tihomirov} et~al., \emph{{Results of experimental investigations of
  cobalt beta decay rate variation}}, {\emph{ArXiv e-prints} (2013) }
  [\href{https://arxiv.org/abs/1304.6885}{{\ttfamily arXiv: 1304.6885}}].

\bibitem{Jenkins2012B}
J.~H. {Jenkins}, E.~{Fischbach}, D.~{Javorsek}, II, R.~H. {Lee} and P.~A.
  {Sturrock}, \emph{{Concerning the Time Dependence of the Decay Rate of
  $^{137}\mathrm{Cs}$}}, {\emph{ArXiv e-prints} (2012) }
  [\href{https://arxiv.org/abs/1211.2138}{{\ttfamily arXiv: 1211.2138}}].

\bibitem{O'Keefe:2012}
D.~{O'Keefe}, B.~L. {Morreale}, R.~H. {Lee}, J.~B. {Buncher}, J.~H. {Jenkins},
  E.~{Fischbach} et~al., \emph{{Spectral content of $^{22}$Na/$^{44}$Ti decay
  data: implications for a solar influence}}, {\emph{APSS} {\bfseries 344}
  (2013) 297} [\href{https://arxiv.org/abs/1212.2198}{{\ttfamily arXiv: 1212.2198}}].

\bibitem{Norman2009}
E.~B. Norman, E.~Browne, H.~A. Shugart, T.~H. Joshi and R.~B. Firestone,
  \emph{{Evidence against correlations between nuclear decay rates and
  Earth-Sun distance}}, {\emph{Astropart.Phys.}
  {\bfseries 31} (2009) 135} [\href{https://arxiv.org/abs/0810.3265}{{\ttfamily arXiv:
  0810.3265}}].

\bibitem{Bernabei2008}
{\scshape DAMA} collaboration, R.~Bernabei et~al., \emph{First
  results from {DAMA/LIBRA} and the combined results with {DAMA/NaI}}, {\emph{Eur.Phys.J.}
  {\bfseries C56} (2008) 333}
  [\href{https://arxiv.org/abs/0804.2741}{{\ttfamily arXiv: 0804.2741}}].

\bibitem{Bernabei2005}
{\scshape DAMA} collaboration, R.~Bernabei et~al., \emph{{Dark
  matter particles in the {Galactic} halo: {Results} and implications from
  {DAMA/NaI}}}, {\emph{Int. J. Mod. Phys.} {\bfseries D13} (2004) 2127}
  [\href{https://arxiv.org/abs/astro-ph/0501412}{{\ttfamily
  astro-ph/0501412}}].

\bibitem{Bernabei2013}
{\scshape DAMA} collaboration, R.~Bernabei et~al., \emph{Dark
  matter investigation by {DAMA} at {Gran Sasso}},
  [\href{https://arxiv.org/abs/1306.1411}{{\ttfamily arXiv: 1306.1411}}].

\bibitem{Agnese:2014}
{\scshape SuperCDMS} collaboration, R.~Agnese et~al., \emph{{Search for
  Low-Mass Weakly Interacting Massive Particles with SuperCDMS}}, {\emph{Phys. Rev. Lett.}
  {\bfseries 112} (2014) 241302}
  [\href{https://arxiv.org/abs/1402.7137}{{\ttfamily arXiv:1402.7137}}].

\bibitem{Aprile2015}
{\scshape XENON100} collaboration, E.~Aprile et~al., \emph{{Exclusion of
  Leptophilic Dark Matter Models using XENON100 Electronic Recoil Data}}, {\emph{Science} {\bfseries 349}
  (2015) 851} [\href{https://arxiv.org/abs/1507.07747}{{\ttfamily arXiv:
  1507.07747}}].

\bibitem{Akerib2017}
{\scshape LUX} collaboration, D.~S. {Akerib} et~al., \emph{{Results from a
  Search for Dark Matter in the Complete LUX Exposure}},
{\emph{Physical Review Letters} {\bfseries 118} (2017) 021303}
  [\href{https://arxiv.org/abs/1608.07648}{{\ttfamily arXiv: 1608.07648}}].

\bibitem{Amole2017}
{\scshape PICO} collaboration, C.~{Amole} et~al., \emph{{Dark Matter Search
  Results from the {PICO -60 $C_{3}$F$_{8}$} Bubble Chamber}}, {\emph{Physical Review Letters} {\bfseries 118} (2017) 251301}
  [\href{https://arxiv.org/abs/1702.07666}{{\ttfamily arXiv: 1702.07666}}].

\bibitem{Aprile:2017B}
{\scshape XENON1T} collaboration, E.~{Aprile} et~al., \emph{{First Dark Matter
  Search Results from the XENON1T Experiment}}, {\emph{ArXiv e-prints} (2017) }
  [\href{https://arxiv.org/abs/1705.06655}{{\ttfamily arXiv: 1705.06655}}].

\bibitem{Pradler:2012A}
J.~Pradler, B.~Singh and I.~Yavin, \emph{{On an unverified nuclear decay and
  its role in the DAMA experiment}},
  {\emph{Phys. Lett.} {\bfseries B720} (2013) 399}
  [\href{https://arxiv.org/abs/1210.5501}{{\ttfamily arXiv: 1210.5501}}].

\bibitem{Pradler:2012B}
J.~Pradler and I.~Yavin, \emph{{Addendum to "On an unverified nuclear decay and
  its role in the DAMA experiment''}}, {\emph{Phys. Lett.}
  {\bfseries B723} (2013) 168}
  [\href{https://arxiv.org/abs/1210.7548}{{\ttfamily arXiv: 1210.7548}}].

\bibitem{Bellotti2012}
E.~Bellotti, C.~Broggini, G.~D. Carlo, M.~Laubenstein and R.~Menegazzo,
  \emph{{Search for the time dependence of the $^{137}\mathrm{Cs}$ decay
  constant}}, {\emph{Phys.Lett.} {\bfseries B710} (2012) 114}
  [\href{https://arxiv.org/abs/1202.3662}{{\ttfamily arXiv: 1202.3662}}].

\bibitem{Baurov2001}
Y.~A. Baurov, A.~A. Konradov, V.~F. Kushniruk, E.~A. Kuznetsov, Y.~G. Sobolev,
  Y.~V. Ryabov et~al., \emph{Experimental investigations of changes in
  $\beta$-decay rate of {$^{60}$Co} and {$^{137}$Cs}}, {\emph{Modern Physics Letters
  A} {\bfseries 16} (2001) 2089} [\href{https://doi.org/10.1142/S0217732301005187}{\ttfamily DOI: 10.1142/S0217732301005187}].

\bibitem{Fischbach2009}
E.~Fischbach, J.~B. Buncher, J.~T. Gruenwald, J.~H. Jenkins, D.~E. Krause,
  J.~J. Mattes et~al., \emph{Time-dependent nuclear decay parameters: New
  evidence for new forces?}, {\emph{Space Science Reviews}
  {\bfseries 145} (2009) 285} [\href{https://doi.org/10.1007/s11214-009-9518-5}{\ttfamily DOI: 10.1007/s11214-009-9518-5}].

\bibitem{Jenkins2008}
J.~H. Jenkins and E.~Fischbach, \emph{{Perturbation of Nuclear Decay Rates
  During the Solar Flare of 13 December 2006}}, {\emph{Astropart.Phys.}
  {\bfseries 31} (2009) 407} [\href{https://arxiv.org/abs/0808.3156}{{\ttfamily arXiv:
  0808.3156}}].

\bibitem{Bellotti2013}
E.~Bellotti, C.~Broggini, G.~D. Carlo, M.~Laubenstein and R.~Menegazzo,
  \emph{{Search for correlations between solar flares and decay rate of
  radioactive nuclei}}, {\emph{Phys.Lett.}
  {\bfseries B720} (2013) 116}
  [\href{https://arxiv.org/abs/1302.0970}{{\ttfamily arXiv: 1302.0970}}].

\bibitem{Schrader2016}
H.~Schrader, \emph{Seasonal variations of decay rate measurement data and their
  interpretation}, {\emph{Applied Radiation and Isotopes} {\bfseries 114} (2016) 202 }
  [\href{https://doi.org/http://dx.doi.org/10.1016/j.apradiso.2016.05.001}{\ttfamily DOI: 10.1016/j.apradiso.2016.05.001}].

\bibitem{Pomme2017alpha}
S.~Pomm\'{e}, H.~Stroh, J.~Paepen, R.~V. Ammel, M.~Marouli, T.~Altzitzoglou
  et~al., \emph{On decay constants and orbital distance to the sun --- part
  {I}: alpha decay}, {\emph{Metrologia} {\bfseries 54} (2017) 1}.

\bibitem{Pomme2017betaec}
S.~Pomm\'{e}, H.~Stroh, J.~Paepen, R.~V. Ammel, M.~Marouli, T.~Altzitzoglou
  et~al., \emph{On decay constants and orbital distance to the sun --- part
  {III}: beta plus and electron capture decay}, {\emph{Metrologia} {\bfseries
  54} (2017) 36}.

\bibitem{Chen:2011}
J.~Chen, B.~Singh and J.~A. Cameron, \emph{Nuclear data sheets for a= 44},
  {\emph{Nuclear Data Sheets} {\bfseries 112} (2011) 2357}.

\bibitem{dong2014}
Y.~Dong and H.~Junde, \emph{Nuclear data sheets for a= 54}, {\emph{Nuclear Data
  Sheets} {\bfseries 121} (2014) 1}.

\bibitem{Tuli:2003}
J.~K. Tuli, \emph{Nuclear data sheets for a= 60}, {\emph{Nuclear Data Sheets}
  {\bfseries 100} (2003) 347}.

\bibitem{Browne:2007}
E.~Browne and J.~K. Tuli, \emph{Nuclear data sheets for a= 137}, {\emph{Nuclear
  Data Sheets} {\bfseries 108} (2007) 2173}.

\bibitem{Agostinelli:2002}
{\scshape GEANT4} collaboration, S.~Agostinelli et~al., \emph{{GEANT4: A
  Simulation toolkit}}, {\emph{Nucl. Instrum. Meth.} {\bfseries A506} (2003) 250} 
  [\href{https://doi.org/10.1016/S0168-9002(03)01368-8}{\ttfamily DOI: 10.1016/S0168-9002(03)01368-8}].

\bibitem{knoll2000}
G.~Knoll, \emph{Radiation Detection and Measurement}. Wiley, 2000.

\bibitem{Engelkemeir:1956}
D.~Engelkemeir, \emph{{Nonlinear Response of NaI(Tl) to Photons}},
{\emph{Review of Scientific Instruments} {\bfseries 27} (1956) 589}
  [\href{http://dx.doi.org/10.1063/1.1715643}{{\ttfamily DOI: 10.1063/1.1715643}}].

\bibitem{Brun1997}
R.~{Brun} and F.~{Rademakers}, \emph{{ROOT --- An object oriented data analysis
  framework}}, {\emph{Nuclear Instruments and Methods in Physics Research A} {\bfseries 389} (1997) 81}
  [\href{https://doi.org/10.1016/S0168-9002(97)00048-X}{\ttfamily DOI: 10.1016/S0168-9002(97)00048-X}].

\bibitem{Agilent:2012}
\emph{Laser and Optics Users Manual}, ch.~17. Material Expansion Coefficients.
\newblock Agilent Technologies, 2012.

\bibitem{Towers2013}
S.~Towers, \emph{Improving the control of systematic uncertainties in precision
  measurements of radionuclide half-life}, {\emph{Applied
  Radiation and Isotopes} {\bfseries 77} (2013) 110 } 
  [\href{https://doi.org/https://doi.org/10.1016/j.apradiso.2013.03.003}{\ttfamily DOI: 10.1016/j.apradiso.2013.03.003}].

\bibitem{Jenkins2009}
J.~H. {Jenkins}, E.~{Fischbach}, J.~B. {Buncher}, J.~T. {Gruenwald}, D.~E.
  {Krause} and J.~J. {Mattes}, \emph{{Evidence of correlations between nuclear
  decay rates and Earth-Sun distance}},
{\emph{Astroparticle Physics} {\bfseries 32} (2009) 42}
  [\href{https://arxiv.org/abs/0808.3283}{{\ttfamily arXiv: 0808.3283}}].

\end{thebibliography}

\providecommand{\href}[2]{#2}\begingroup\raggedright\endgroup


\end{document}